**Defining and Estimating Outcomes Directly Averted by a Vaccination Program when Rollout Occurs Over Time**

**Authors:** Katherine M Jia[1§], Christopher B Boyer[2,3], Alyssa Bilinski[4,5], Marc Lipsitch[1,6]

**Affiliations:**

[1] Center for Communicable Disease Dynamics, Department of Epidemiology, Harvard T.H. Chan School of Public Health, Boston, Massachusetts, USA

[2] Department of Quantitative Health Sciences, Cleveland Clinic Research, Cleveland, Ohio, USA

[3] Department of Medicine, Cleveland Clinic Lerner College of Medicine of Case Western Reserve University, Cleveland, Ohio, USA

[4] Department of Health Services, Policy, and Practice, Brown University School of Public Health, Providence, Rhode Island, USA

[5] Department of Biostatistics, Brown University School of Public Health, Providence, Rhode Island, USA

[6] Department of Immunology and Infectious Diseases, Harvard T.H. Chan School of Public Health, Boston, Massachusetts, USA

[§] **Correspondence:** Katherine M Jia, Ph.D., Center for Communicable Disease Dynamics, Department of Epidemiology, Harvard T.H. Chan School of Public Health, 677 Huntington Avenue, Boston, MA 02115, USA. +1 (617) 432-7955. (kjia@hsph.harvard.edu).

**Sources of funding**: K.M.J and M.L. were supported by a subaward from CDC cooperative agreement to Carnegie-Mellon University (Grant number: U01 IP001121). A.B. was supported by National Institutes of General Medical Studies (Grant number: 1R35GM155224).

**Abstract (248 words)**

During the COVID-19 pandemic, estimating the total deaths averted by vaccination has been of great public health interest. Instead of estimating total deaths averted by vaccination among both vaccinated and unvaccinated individuals, some studies empirically estimated only "directly averted" deaths among vaccinated individuals, typically suggesting that vaccines prevented more deaths overall than directly due to the indirect effect. Here, we define the causal estimand to quantify outcomes "directly averted" by vaccination—i.e., the impact of vaccination for vaccinated individuals, holding vaccination coverage fixed—for vaccination at multiple time points, and show that this estimand is a lower bound on the total outcomes averted when the indirect effect is non-negative. We develop an unbiased estimator for the causal estimand in a one-stage randomized controlled trial (RCT) and explore the bias of a popular "hazard difference" estimator frequently used in empirical studies. We show that even in an RCT, the hazard difference estimator is biased if vaccination has a non-null effect, as it fails to incorporate the greater depletion of susceptibles among the unvaccinated individuals. In simulations, the overestimation is small for averted deaths when infection-fatality rate is low, as for many important pathogens. However, the overestimation can be large for averted infections given a high basic reproduction number. Additionally, we define and compare estimand and estimators for avertible outcomes (i.e., outcomes that could have been averted by vaccination, but were not due to failure to vaccinate). Future studies can explore the identifiability of the causal estimand in observational settings.

## 1 Introduction

During the COVID-19 pandemic, determining the total number of infections (or deaths) averted by vaccination has been of great public health interest.[1–10] Researchers are interested in how many infections (or deaths) have been averted overall by COVID-19 vaccine rollout programs, compared to the counterfactual of no vaccination for anyone. However, in the presence of indirect effects, the key challenge is that we may not observe a comparable population that is unvaccinated throughout. Rather than estimating the total number of outcomes averted among both vaccinated and unvaccinated individuals, some empirical studies[4–8] have instead estimated outcomes "directly averted" among vaccinated individuals by conditioning on the actual vaccination coverage in the rest of the population. The estimand and estimation procedures used in selected examples of studies are summarized in **eAppendix 1**. Typically, studies that estimated outcomes directly averted among vaccinated individuals computed daily or weekly hazards of death among all vaccinated and not-yet-vaccinated individuals, calculated the difference in hazards, multiplied this difference with the number of vaccinated survivors, and summed the results across time. We refer to this method as the "hazard difference estimator." Empirical analyses of this type commonly assumed that such directly averted outcomes are a lower bound on the total averted outcomes due to the indirect effect in reducing transmission.[4–6]

This study is motivated by two research gaps from the numerous empirical analyses[4–8,11,12] that estimated directly averted (or avertible) outcomes among vaccinated (or unvaccinated) individuals under a vaccine rollout. First, the causal estimand for directly averted outcomes has not been precisely defined as a mathematical quantity for vaccination at multiple time points under interference. Second, the popular hazard difference estimator used by these analyses has not been evaluated. Therefore, we first propose the casual estimand and its unbiased estimator based on a one-stage randomized controlled trial (RCT). We use the causal estimand to formalize the lower bound assumption and identify the condition under which a vaccine rollout program has averted more outcomes overall than directly among the vaccinated individuals. Last, we evaluate the bias of the hazard difference estimator relative to the causal estimand. This paper is an important extension to our previous study[13] which develops estimands for quantifying averted outcomes for vaccination at a single time point, as vaccination almost always occurs over time in reality.

Section 2 of this paper describes the setup and notation. Section 3 defines the causal estimand for quantifying directly averted outcomes among vaccinated individuals, as well as its unbiased estimator and the hazard difference estimator. Section 4 examines the bias of the hazard difference estimator both analytically and through simulations.

**2 Setup and notation**

Hudgens and Halloran[14] defined four *effect* estimands (namely the direct, indirect, total, and overall effects) for vaccination at a single time point in a two-stage randomized trial, which reduces to a one-stage RCT when the study population consists of a single group. In reality, two-stage RCTs are rarely conducted or justified; for our purpose of estimating directly averted outcomes, we define our estimand and estimators based on a one-stage RCT, with notation closely aligned to that of Hudgens and Halloran.[14]

Consider a one-stage RCT that consists of $N$ individuals indexed by $j = 1, \dots, N$ with a large $N$.[14] Consider $q + 2$ evenly spaced measurement intervals for $q \in \mathbb{N}$. Let $l \in \{0, \dots, q + 1\}$ denote each interval with baseline measurements taken in interval 0, and $q + 1$ representing the end of follow-up.[15] Let $X_j$ denote the assigned vaccination time, where $x_j \in \{0, \dots, q + 1\}$ are possible realizations of $X_j$.[16] Vaccination occurs *at the beginning* of each interval, whereas $x_j = q + 1$ denotes unvaccinated throughout. Let $\mathbf{X} = (X_1, \dots, X_N)$ denote the vaccination times individuals were assigned. Let $\mathbf{x}$ denote a possible realization of $\mathbf{X}$. Let $\mathcal{X}(N)$ denote the set of all possible $(q+1)^N$ vaccination time allocations for the group, for which $\mathbf{x} \in \mathcal{X}(N)$. Throughout we assume perfect compliance (i.e., assignment to a particular vaccination time is equivalent to receipt of vaccination at that time if the person is still alive), no loss-to-follow-up, and no measurement error. In an ideal one-stage RCT, these assumptions are expected to hold.

Here, interference is assumed—that is, the potential outcome for any individual depend on vaccination assignments of every other individual in the group.[14] Let $Y_{q+1,j}(\mathbf{x}) \in \{0,1\}$ be a *cumulative* indicator of experiencing the outcome (e.g., death) *before the beginning* of interval $q+1$ for individual $j$ had the group followed the vaccination schedule $\mathbf{x} \in \mathcal{X}(N)$. By convention, $Y_{0,j} \equiv 0$.[17]

Let $\boldsymbol{\rho} = \{q+1, d; \rho_0, \dots, \rho_q\}$ denote parameterizations that govern the distribution of $\mathbf{X}$, where $q+1$ is the number of potential vaccination times, $d$ is the number of days of each interval, $\rho_x$ is the proportion of individuals assigned to $x \in \{0, \dots, q+1\}$ such that $\sum_{x=0}^{q+1} \rho_x = 1$. We assume $\boldsymbol{\rho}$ is a *mixed individual assignment strategy*,[14,18] as defined in **eAppendix 2**. In words, $\boldsymbol{\rho}$ randomly assigns $\rho_0 \times 100\%$ of the individuals to receive vaccination at baseline, $\rho_1 \times 100\%$ to receive vaccination at the beginning of interval 1, and so on, with $\rho_{q+1} \times 100\%$ to remain unvaccinated throughout.[18] Let $\boldsymbol{\phi} = \{q+1, d; \mathbf{0}\}$ denote no vaccination. To quantify vaccine-averted outcomes, our goal is to assess the impact of some vaccination strategy $\boldsymbol{\rho}$ compared to $\boldsymbol{\phi}$.

At baseline, individuals are randomly assigned to $\mathbf{X}$ conditional on a mixed individual assignment strategy (i.e., fixed proportions of individuals [e.g., 20%, 30%, 50%] were assigned to receive vaccination at specific times [e.g., Day 0, Day 60, Day 120], respectively; see definition in **eAppendix 2**). Note that random assignments $\mathbf{X}$ occurs only at baseline, even though individuals receive vaccination at different times. This study design is referred to as a "one-stage" RCT because randomization takes place only once, in contrast to the "two-stage" RCT described by Hudgens and Halloran,[14] in which groups are first randomized to different strategies and then individuals within each group are randomized to vaccination conditional on their group's assigned strategy. Let $\bar{Y}_{q+1}(x;\boldsymbol{\rho})$ denote the group average potential outcome as defined in **eAppendix 2**, which is equivalent to population average potential outcome[14] because there is only one group. Let $\Delta\bar{Y}_{l+1}(x;\boldsymbol{\rho}) = \bar{Y}_{l+1}(x;\boldsymbol{\rho}) - \bar{Y}_{l}(x;\boldsymbol{\rho})$ for $l \in \{0,\ldots,q\}$ be the difference in $\bar{Y}$ between intervals $l$ and $l+1$.[19]

## 3 Causal estimand and estimators for quantifying outcomes directly averted by vaccination

### 3.1 Causal estimand

In our prior work with vaccination at a single time point,[13] we defined the "direct impact" estimand to quantify outcomes directly averted among vaccinated individuals as the number of vaccinated individuals multiplied by the direct effect (DE)[14] for vaccination at the baseline time point. In notation, the number of outcomes directly averted by $\boldsymbol{\rho} = \{1,d;\rho_0\}$ (i.e, vaccination of $\rho_0$ at the beginning of a single interval, with interval duration $d$), compared to no vaccination $\boldsymbol{\phi} = \{1,d;0\}$, is:

$$\begin{aligned}\delta_1^D(\boldsymbol{\phi},\boldsymbol{\rho}) &= N\rho_0\big(\bar{Y}_1(1;\boldsymbol{\rho}) - \bar{Y}_1(0;\boldsymbol{\rho})\big)\\ &= N\rho_0 DE_1((1,0);\boldsymbol{\rho}). \qquad (1)\end{aligned}$$

When vaccination occurs at multiple time points, there could be multiple versions of direct effects. A direct effect can be a contrast between non-vaccination and some vaccination time $x' \in \{0,\ldots,q\}$, conditional on $\boldsymbol{\rho}$ (e.g., comparing individuals unvaccinated versus vaccinated at interval 0, or unvaccinated versus vaccinated at interval 1). In notation, the direct effect comparing probability of having developed the outcome by the beginning of interval $q+1$ for an individual unvaccinated throughout versus assigned to $x'$ when the group follows strategy $\boldsymbol{\rho} = \{q + 1, d; \rho_0, \dots, \rho_q\}$ is:

$$DE_{q+1}((q+1, x'); \boldsymbol{\rho}) = \overline{Y}_{q+1}(q+1; \boldsymbol{\rho}) - \overline{Y}_{q+1}(x'; \boldsymbol{\rho}). \quad (2)$$

Now, extending the estimand in equation (1) for vaccination at two time points (i.e., $x' = 0$ or $x' = 1$, as compared to unvaccinated $x = 2$), we consider the direct effect of $x' \in \{0,1\}$ compared to unvaccinated throughout, weighted by the number of individuals assigned with $x'$, and summed across $x' \in \{0,1\}$. In notation, the number of outcomes directly averted by $\boldsymbol{\rho} = \{2, d; \rho_0, \rho_1\}$, compared to no vaccination $\boldsymbol{\phi} = \{2, d; \mathbf{0}\}$, is:

*Definition 1 (Causal estimand for directly averted outcomes for vaccination at two time points)*

$$\delta_2^D(\boldsymbol{\phi}, \boldsymbol{\rho}) = N \cdot \{\rho_0 \cdot [\overline{Y}_2(2; \boldsymbol{\rho}) - \overline{Y}_2(0; \boldsymbol{\rho})] + \rho_1 \cdot [\overline{Y}_2(2; \boldsymbol{\rho}) - \overline{Y}_2(1; \boldsymbol{\rho})]\} \quad (3)$$

$$= N \cdot \left\{\rho_0 \cdot DE_2\big((2,0); \boldsymbol{\rho}\big) + \rho_1 \cdot DE_2\big((2,1); \boldsymbol{\rho}\big)\right\}.$$

**eAppendix 3** extends $\delta_2^D(\boldsymbol{\phi}, \boldsymbol{\rho})$ to an arbitrary number of vaccination times. **eAppendix 4** shows that $\delta_{q+1}^D(\boldsymbol{\phi}, \boldsymbol{\rho})$ is the lower bound on outcomes averted among both vaccinated and unvaccinated individuals when the indirect effect is non-negative. However, the indirect effect can be negative when transmission or fatality parameters vary over time, as we showed earlier for vaccination at a single time point.[13] Therefore, in the more general case of vaccination at multiple time points, the indirect effect is not guaranteed to be non-negative under many realistic scenarios.

**3.2 Unbiased estimator**

Define $\hat{Y}_2(x; \boldsymbol{\rho}) = \frac{\sum_{j=1}^{N} Y_{2,j}(\mathbf{X}) I[X_j = x]}{\sum_{j=1}^{N} I[X_j = x]}$ for $x \in \{0,1,2\}$. That is, $\hat{Y}_2(x; \boldsymbol{\rho})$ is the cumulative incidence by the beginning of interval 2 for individuals assigned to $x$ under strategy $\boldsymbol{\rho}$. Then the estimator for $\delta_2^D(\boldsymbol{\phi}, \boldsymbol{\rho})$ in equation (3) is:

$$\widehat{\delta_2}^D(\boldsymbol{\phi}, \boldsymbol{\rho}) = N \cdot \left\{\rho_0 \cdot \left[\hat{Y}_2(2; \boldsymbol{\rho}) - \hat{Y}_2(0; \boldsymbol{\rho})\right] + \rho_1 \cdot \left[\hat{Y}_2(2; \boldsymbol{\rho}) - \hat{Y}_2(1; \boldsymbol{\rho})\right]\right\}. \quad (4)$$

**eAppendix 5** proves that $\widehat{\delta_{q+1}}^{D}(\boldsymbol{\phi}, \boldsymbol{\rho})$ is an unbiased estimator for $\delta_{q+1}^{D}(\boldsymbol{\phi}, \boldsymbol{\rho})$ for $q \in \mathbb{N}$ in a one-stage RCT under mixed assignment strategy $\boldsymbol{\rho}$. Note that $\rho_x$ for $x \in \{0,1, \dots, q+1\}$ is known and fixed under a mixed assignment strategy $\boldsymbol{\rho}$ in an RCT, although it must be estimated in an observational setting.

### 3.3 Hazard difference estimator

As summarized in the literature review (**eAppendix 1**), recent empirical studies[4–8] often used what we refer to as the *hazard difference* estimator, which relies on the number at risk and the number of new cases among individuals vaccinated (and not-yet-vaccinated) by a given time to quantify outcomes directly averted among vaccinated individuals.

Let $\Delta\hat{Y}_{l+1}(x; \boldsymbol{\rho}) = \hat{Y}_{l+1}(x; \boldsymbol{\rho}) - \hat{Y}_{l}(x; \boldsymbol{\rho})$ for $l \in \{0, \dots, q\}$. Equation (5) defines the hazard difference estimator for two vaccination times (see **eAppendix 6** for extension to an arbitrary number of vaccination times).

*Definition 2 (Hazard difference estimator for directly averted outcomes for vaccination at two time points)*

$$\widehat{\delta_{2}}^{D*}(\boldsymbol{\phi}, \boldsymbol{\rho}) = \hat{N}_0^{v}(\boldsymbol{\rho})\left(\hat{h}_1^{u}(\boldsymbol{\rho}) - \hat{h}_1^{v}(\boldsymbol{\rho})\right) + \hat{N}_1^{v}(\boldsymbol{\rho})\left(\hat{h}_2^{u}(\boldsymbol{\rho}) - \hat{h}_2^{v}(\boldsymbol{\rho})\right) \quad (5)$$

where $\hat{h}_1^{v}(\boldsymbol{\rho}) = \Delta\hat{Y}_1(0; \boldsymbol{\rho})$ is the incidence among vaccinated individuals by the beginning of interval 1, $\hat{h}_1^{u}(\boldsymbol{\rho}) = \frac{\rho_1 \Delta\hat{Y}_1(1;\boldsymbol{\rho}) + \rho_2 \Delta\hat{Y}_1(2;\boldsymbol{\rho})}{\rho_1 + \rho_2}$ is the incidence among not-yet-vaccinated individuals by the beginning of interval 1 (Note this quantity is the combination of two distinct groups—individuals assigned to $x = 1$ and $x = 2$), $\hat{N}_0^{v}(\boldsymbol{\rho}) = N\rho_0$ is the number survived by the beginning of interval 0 among individuals assigned to $x = 0$, $\hat{N}_1^{v}(\boldsymbol{\rho}) = N\left[\rho_0\left(1 - \hat{Y}_1(0; \boldsymbol{\rho})\right) + \rho_1\left(1 - \hat{Y}_1(1; \boldsymbol{\rho})\right)\right]$ is the combined number survived by the beginning of interval 1 among individuals assigned to $x = 0$ and $x = 1$, $\widehat{h}_2^v(\boldsymbol{\rho}) = \frac{\rho_0 \Delta\widehat{Y}_2(0;\boldsymbol{\rho}) + \rho_1 \Delta\widehat{Y}_2(1;\boldsymbol{\rho})}{\rho_0\left(1-\widehat{Y}_1(0;\boldsymbol{\rho})\right) + \rho_1\left(1-\widehat{Y}_1(1;\boldsymbol{\rho})\right)}$ is the hazard among vaccinated individuals by the beginning of interval 2, $\widehat{h}_2^u(\boldsymbol{\rho}) = \frac{\Delta\widehat{Y}_2(2;\boldsymbol{\rho})}{1-\widehat{Y}_1(2;\boldsymbol{\rho})}$ is the hazard among not-yet-vaccinated individuals by the beginning of interval 2.

To give intuition for the unbiased estimator and the hazard difference estimator, consider the following example (**Figure 1**). Suppose we observe an ideal trial randomizing 6 individuals into three arms according to the strategy $\boldsymbol{\rho}^* = \{2,60; \frac{1}{3}, \frac{1}{3}\}$. One-third of the individuals are randomly assigned to vaccination at baseline ($x = 0$), one-third to vaccination at the beginning of interval 1 ($x = 1$), and the remaining to no vaccination throughout ($x = 2$). As illustrated in **Figure 1A**, the unbiased estimator gives $\widehat{\delta_2}^D(\boldsymbol{\phi}, \boldsymbol{\rho}^*) = 6 \cdot \left\{\rho_0 \cdot \left[\widehat{Y}_2(2;\boldsymbol{\rho}^*) - \widehat{Y}_2(0;\boldsymbol{\rho}^*)\right] + \rho_1 \cdot \left[\widehat{Y}_2(2;\boldsymbol{\rho}^*) - \widehat{Y}_2(1\,;\boldsymbol{\rho}^*)\right]\right\} = 6 \cdot \left[\frac{1}{3} \cdot (1-0) + \frac{1}{3} \cdot \left(1 - \frac{1}{2}\right)\right] = 3$ . As illustrated in **Figure 1B**, the hazard difference estimator gives $\widehat{\delta_2}^{D*}(\boldsymbol{\phi}, \boldsymbol{\rho}^*) = \widehat{N}_0^v(\boldsymbol{\rho}^*)\left(\widehat{h}_1^u(\boldsymbol{\rho}^*) - \widehat{h}_1^v(\boldsymbol{\rho}^*)\right) + \widehat{N}_1^v(\boldsymbol{\rho}^*)\left(\widehat{h}_2^u(\boldsymbol{\rho}^*) - \widehat{h}_2^v(\boldsymbol{\rho}^*)\right) = 2 \cdot \left(\frac{1}{2} - 0\right) + 3 \cdot (1-0) = 4.$

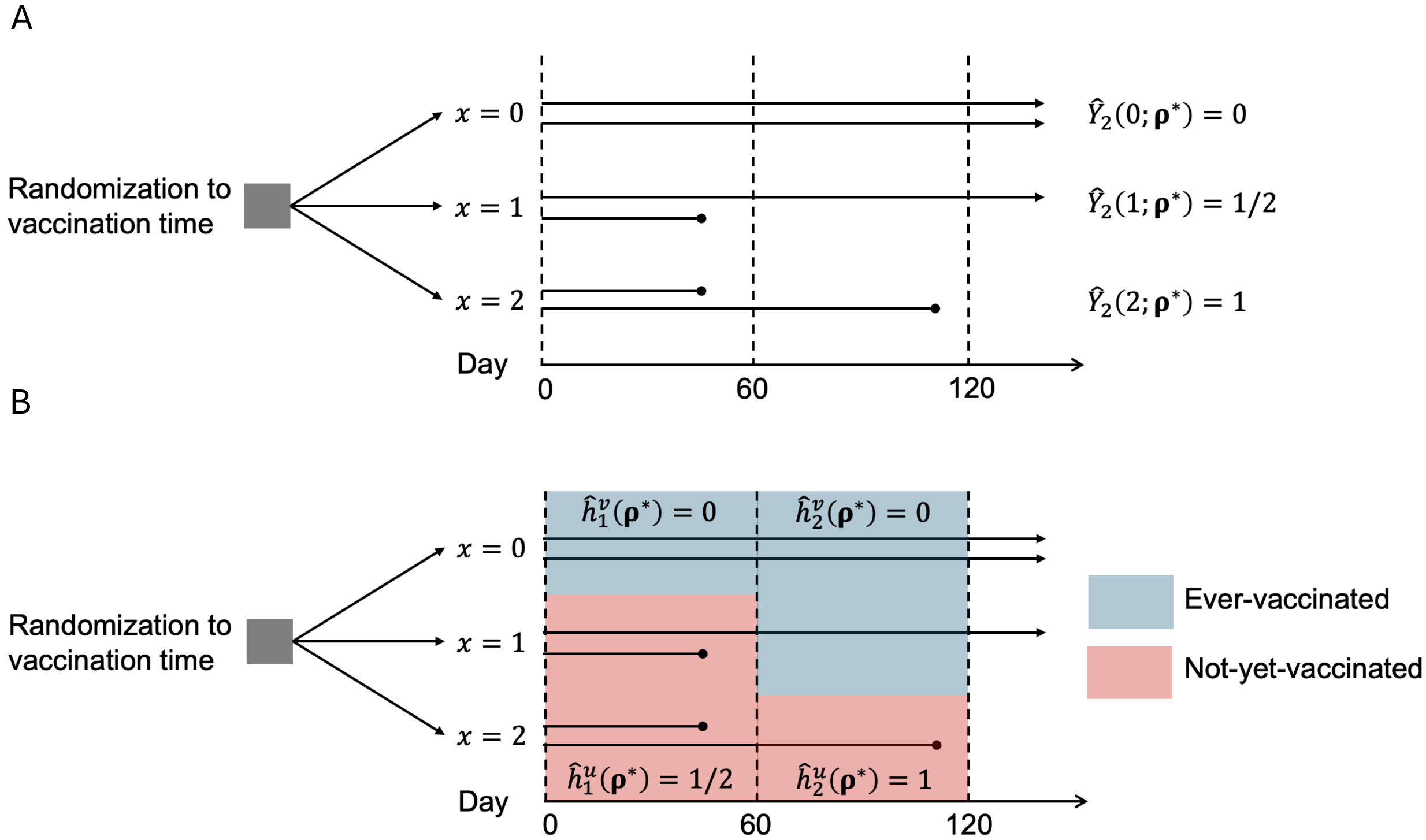

**FIGURE 1** | Schematic representation of an ideal randomized controlled trial with 6 individuals (horizontal lines) under the strategy $\boldsymbol{\rho}^* = \{2,60;\frac{1}{3},\frac{1}{3}\}$. One-third of individuals are randomly assigned to vaccination at baseline ($x = 0$), one third to vaccination at the beginning of interval 1 ($x = 1$), and the remaining to no vaccination throughout ($x = 2$). The dashed vertical lines represent the beginning of each interval. (A) The unbiased estimator considers the cumulative incidence in each arm. Note $\hat{Y}_l(x; \boldsymbol{\rho}^*)$ is the cumulative incidence by the beginning of interval $l$ among individuals assigned to vaccination time $x$ under strategy $\boldsymbol{\rho}^*$. (B) Hazard difference estimator considers the hazards and survival among vaccinated and not-yet-vaccinated individuals. Note $\hat{h}_{l+1}^v(\cdot)$ (or $\hat{h}_{l+1}^u(\cdot)$) is the hazard for vaccinated (or not-yet-vaccinated) individuals from interval $l$ to $l + 1$.

**4 Bias of the hazard difference estimator for the causal estimand**

Now, we use analytical and simulation approaches to examine the bias of $\widehat{\delta_2}^{D*}(\boldsymbol{\phi}, \boldsymbol{\rho})$ relative to the causal estimand.

**4.1 Analytic comparison**

First, re-write the causal estimand $\delta_2^D(\boldsymbol{\phi}, \boldsymbol{\rho})$ as follows (See derivation in **eAppendix 3**):

$$
\begin{aligned}
&\delta_2^D(\boldsymbol{\phi}, \boldsymbol{\rho}) \\
&= N \cdot \left\{ \rho_0 \cdot \left( \frac{\rho_1 \Delta \bar{Y}_1(1; \boldsymbol{\rho}) + \rho_2 \Delta \bar{Y}_1(2; \boldsymbol{\rho})}{\rho_1 + \rho_2} - \Delta \bar{Y}_1(0; \boldsymbol{\rho}) \right) \right. \\
&\qquad \left. + \left[ (\rho_0 + \rho_1) \cdot \Delta \bar{Y}_2(2; \boldsymbol{\rho}) - \left( \rho_0 \Delta \bar{Y}_2(0; \boldsymbol{\rho}) + \rho_1 \Delta \bar{Y}_2(1; \boldsymbol{\rho}) \right) \right] \right\} \qquad (6)
\end{aligned}
$$

Then, expand $E\left[\widehat{\delta_2}^{D*}(\boldsymbol{\phi}, \boldsymbol{\rho})\right]$ as follows (See derivation in equation [S13] **eAppendix 6**):

$$
\begin{aligned}
&E\left[\widehat{\delta_2}^{D*}(\boldsymbol{\phi}, \boldsymbol{\rho})\right] = E\left[\widehat{N}_0^v(\boldsymbol{\rho}) \left( \hat{h}_1^u(\boldsymbol{\rho}) - \hat{h}_1^v(\boldsymbol{\rho}) \right) + \widehat{N}_1^v(\boldsymbol{\rho}) \left( \hat{h}_2^u(\boldsymbol{\rho}) - \hat{h}_2^v(\boldsymbol{\rho}) \right) \right] \\
&= N \left\{ \rho_0 \cdot \left( \frac{\rho_1 \Delta \bar{Y}_1(1; \boldsymbol{\rho}) + \rho_2 \Delta \bar{Y}_1(2; \boldsymbol{\rho})}{\rho_1 + \rho_2} - \Delta \bar{Y}_1(0; \boldsymbol{\rho}) \right) \right. \\
&\qquad + E\left[ (\rho_0 + \rho_1) \cdot \frac{\rho_0 \left(1 - \hat{Y}_1(0; \boldsymbol{\rho})\right) + \rho_1 \left(1 - \hat{Y}_1(1; \boldsymbol{\rho})\right)}{(\rho_0 + \rho_1)\left(1 - \hat{Y}_1(2; \boldsymbol{\rho})\right)} \cdot \Delta \hat{Y}_2(2; \boldsymbol{\rho}) \right] \\
&\qquad \left. - \left( \rho_0 \Delta \bar{Y}_2(0; \boldsymbol{\rho}) + \rho_1 \Delta \bar{Y}_2(1; \boldsymbol{\rho}) \right) \right\}. \qquad (7)
\end{aligned}
$$

If $\frac{\rho_0(1-\hat{Y}_1(0;\boldsymbol{\rho}))+\rho_1(1-\hat{Y}_1(1;\boldsymbol{\rho}))}{(\rho_0+\rho_1)(1-\hat{Y}_1(2;\boldsymbol{\rho}))} = 1$, then equation (7) equals (6) (See proof in **eAppendix 6**). That is, $\widehat{\delta_2}^{D*}(\boldsymbol{\phi}, \boldsymbol{\rho})$ is an unbiased estimator under the null (i.e., when the survival is the same between those assigned to no vaccination and those assigned with $x = 0$ or $1$). However, if $\frac{\rho_0(1-\hat{Y}_1(0;\boldsymbol{\rho}))+\rho_1(1-\hat{Y}_1(1;\boldsymbol{\rho}))}{(\rho_0+\rho_1)(1-\hat{Y}_1(2;\boldsymbol{\rho}))} \neq 1$, then $\widehat{\delta_2}^{D*}(\boldsymbol{\phi}, \boldsymbol{\rho})$ is biased relative to the causal estimand, implying that it is biased if vaccination has a non-null effect.

**eAppendix 7** provides an alternative unbiased estimator for quantifying directly averted outcomes with a similar (but not identical) expression and have same data requirement as $\widehat{\delta_2}^{D*}(\boldsymbol{\phi}, \boldsymbol{\rho})$. The alternative unbiased estimator allows estimation of directly averted deaths using data aggregated by vaccination status.

### 4.2 Simulation comparison

#### 4.2.1 Scenarios

We simulate an epidemic with strategy $\boldsymbol{\rho}' = \{2, 60; 0.2, 0.3\}$ under different infection-fatality rate (IFR), vaccine efficacy against infection ($VE_{inf}$), or vaccine efficacy against death given infection ($VE_{death}$). We examine the bias of the hazard difference estimator $\widehat{\delta_2}^{D*}(\boldsymbol{\phi}, \boldsymbol{\rho}')$ relative to the causal estimand $\delta_2^D(\boldsymbol{\phi}, \boldsymbol{\rho}')$, where $\boldsymbol{\phi} = \{2, 60; \mathbf{0}\}$, and identify the conditions under which the bias would be substantial (**Table 1**).

**TABLE 1** | Scenarios for simulations, varied by infection-fatality rate, vaccine efficacy against infection ($VE_{inf}$) and vaccine efficacy against death given infection ($VE_{death}$)

| Scenario | Infection-fatality rate | Vaccine efficacy |
|---|---|---|
| Scenario 1 | 1% | $VE_{inf}$ = 90%; $VE_{death}$ = 0% |
| Scenario 2 | 10% | $VE_{inf}$ = 90%; $VE_{death}$ = 0% |
| Scenario 3 | 100% | $VE_{inf}$ = 90%; $VE_{death}$ = 0% |
| Scenario 4 | 1% | $VE_{inf}$ = 0%; $VE_{death}$ = 90% |
| Scenario 5 | 10% | $VE_{inf}$ = 0%; $VE_{death}$ = 90% |
| Scenario 6 | 100% | $VE_{inf}$ = 0%; $VE_{death}$ = 90% |
| Scenario 7 | 1% | $VE_{inf}$ = 90%; $VE_{death}$ = 90% |
| Scenario 8 | 10% | $VE_{inf}$ = 90%; $VE_{death}$ = 90% |
| Scenario 9 | 100% | $VE_{inf}$ = 90%; $VE_{death}$ = 90% |

In the main text, we explore scenarios with varying IFR, $VE_{inf}$, and $VE_{death}$, including several extreme scenarios for illustrative purposes. For sensitivity analyses, we explore scenarios with varying number of effective contacts ($\beta$), as well as more realistic parameter values for $\beta$, IFR, $VE_{inf}$, and $VE_{death}$, specifically corresponding to seasonal flu, measles, and COVID-19 (wild-type strain).

### 4.2.2 Model

Consider a hypothetical RCT with strategy $\boldsymbol{\rho}' = \{2, 60; 0.2, 0.3\}$, in which interval 0 spans Days 0-59, interval 1 spans Days 60-119, and interval 2 is the post-follow-up period on or after Day 120, such that 20% of individuals can be assigned to vaccination at Day 0, 30% to vaccination at Day 60, or the rest remain unvaccinated throughout. In the susceptible-infected-recovered-death (SIRD) model, individuals are stratified by time of vaccination. Within each stratum, we specify a continuous-time SIRD model. The subscript 0 represents those assigned to receive vaccination in the beginning of Day 0, 1 for those assigned to receive vaccination in the beginning of Day 60, and 2 for the never vaccinated. For those receiving vaccination in the beginning of Day 60, vaccine efficacies against infection ($\theta_1$) and death given infection ($\kappa_1$) are time-varying variables that come in effect on and after Day 60 (**Table S4**). The SIRD model is defined in term of continuous time $t$ as follows:

$$\left.\begin{aligned}
\frac{dS_2(t)}{dt} &= -\lambda(t) \cdot S_2(t) \\
\frac{dS_1(t)}{dt} &= -\theta_1(t) \cdot \lambda(t) \cdot S_1(t) \\
\frac{dS_0(t)}{dt} &= -\theta \cdot \lambda(t) \cdot S_0(t) \\
\frac{dI_2(t)}{dt} &= \lambda(t) \cdot S_2(t) - \gamma \cdot I_2(t) \\
\frac{dI_1(t)}{dt} &= \theta_1(t) \cdot \lambda(t) \cdot S_1(t) - \gamma \cdot I_1(t) \\
\frac{dI_0(t)}{dt} &= \theta \cdot \lambda(t) \cdot S_0(t) - \gamma \cdot I_0(t) \\
\frac{dR_2(t)}{dt} &= (1-\mu) \cdot \gamma \cdot I_2(t) \\
\frac{dR_1(t)}{dt} &= (1-\kappa_1(t) \cdot \mu) \cdot \gamma \cdot I_1(t) \\
\frac{dR_0(t)}{dt} &= (1-\kappa \cdot \mu) \cdot \gamma \cdot I_0(t) \\
\frac{dD_2(t)}{dt} &= \mu \cdot \gamma \cdot I_2(t) \\
\frac{dD_1(t)}{dt} &= \kappa_1(t) \cdot \mu \cdot \gamma \cdot I_1(t) \\
\frac{dD_0(t)}{dt} &= \kappa \cdot \mu \cdot \gamma \cdot I_0(t)
\end{aligned}\right\} (8)$$

where $\gamma =$ recovery rate, $\lambda(t) = \beta \cdot \frac{I_0(t)+I_1(t)+I_2(t)}{N(t)}$ (the hazard rate of infection), $\beta =$ the number of effective contacts made by a typical infectious individual per unit time, $\mu =$ probability of death due to infection, $N(t) =$ sum of all individuals alive at $t$. **eAppendix 8** shows the parameters and initial values used in simulations. For simulation, the model was discretized to day time-steps. Code is available at https://github.com/katjia/vax_rollout_impact.

**4.2.3. Simulations**

**Figure 2** compares the expected value of hazard difference estimator (i.e., $E\left[\widehat{\delta_2}^{D*}(\boldsymbol{\phi}, \boldsymbol{\rho}')\right]$) with the causal estimand (i.e., $\delta_2^D(\boldsymbol{\phi}, \boldsymbol{\rho}')$) under Scenarios outlined in **Table 1**; **eAppendix 9** shows the bias of the hazard difference estimator relative to the causal estimand on the absolute and relative scales.

No infections were averted when $VE_{inf}$ = 0% (**Figure 2A;** Scenarios 4 to 6). In Scenarios where $VE_{inf}$ = 90%, the hazard difference estimator substantially overestimates the averted infections (**Figure 2A**; Scenarios 1 to 3, 7 to 9). This is because given the high reproduction number ($R_0 \approx 3.57$) and a high $VE_{inf}$, susceptibles were preferentially depleted among the unvaccinated individuals quickly, such that by interval 1, individuals with $x = 2$ had lower survival from infection than the average survival among individuals with $x = 0$ and $x = 1$ (**Figure S3**)—that is, $\frac{\rho_0'\left(1-\hat{Y}_1(0;\boldsymbol{\rho}')\right)+\rho_1'\left(1-\hat{Y}_1(1;\boldsymbol{\rho}')\right)}{(\rho_0'+\rho_1')\left(1-\hat{Y}_1(2;\boldsymbol{\rho}')\right)} > 1$ in equation (7). **eAppendix 11** varies the Scenarios by the number of effective contacts ($\beta$) and show that the overestimation is more pronounced the higher $\beta$ is.

The hazard difference estimator substantially overestimates the averted deaths when IFR=100%, while the overestimation is trivial when IFR≤10% (**Figure 2B**). This is due to the similar survival from death between vaccinated and unvaccinated individuals when IFR is low—that is, $\frac{\rho_0'\left(1-\hat{Y}_1(0;\boldsymbol{\rho}')\right)+\rho_1'\left(1-\hat{Y}_1(1;\boldsymbol{\rho}')\right)}{(\rho_0'+\rho_1')\left(1-\hat{Y}_1(2;\boldsymbol{\rho}')\right)} \approx 1.$

**eAppendix 6** repeats the same analyses for vaccination at additional time points and over a longer period, which shows that averted infections are even more severely overestimated compared with vaccination at two time points. Consistent with the main-text finding, the hazard difference estimator sightly overestimates averted deaths when IFR≤10%. In **eAppendix 12**, we consider more realistic parameter values for seasonal flu, measles, and COVID-19 (wild-type strain). As with the main analyses, $\widehat{\delta_2}^{D*}(\boldsymbol{\phi}, \boldsymbol{\rho}')$ substantially overestimates averted infections for measles due to high basic reproduction number ($R_0 = 18$) and high $VE_{inf}$. It also slightly overestimates averted infections for seasonal flu and COVID-19 (wild-type), given a low $R_0$ (i.e., ≤2.2) and $VE_{inf}$ > 0.

The overestimation of averted deaths is trivial due to the low IFR (≤ 3% for all pathogens) (**Figure S5**).

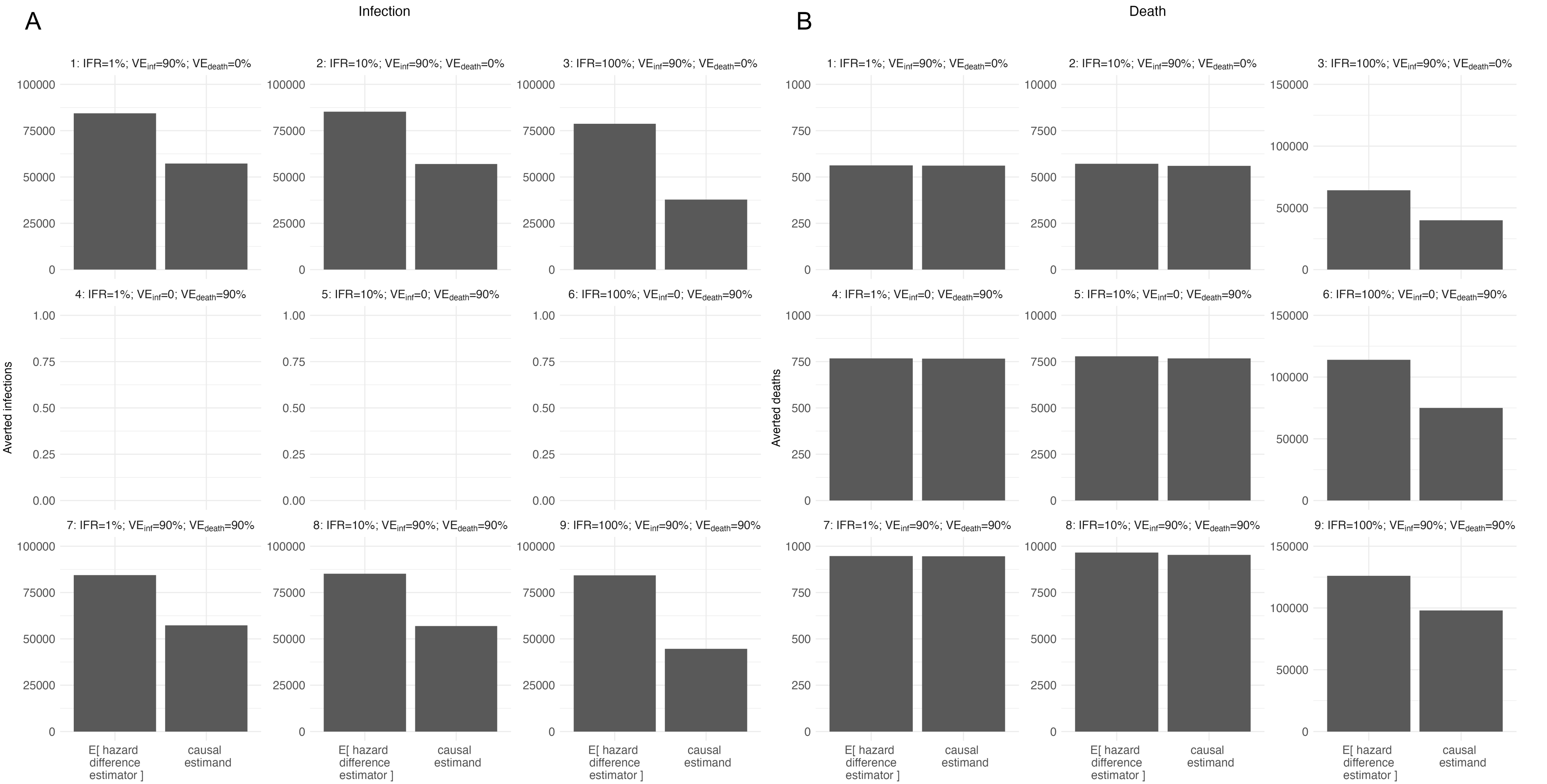

**FIGURE 2** | Infections (A) and deaths (B) directly averted by strategy $\boldsymbol{\rho}' = \{2,60; 0.2,0.3\}$ under different scenarios varied by infection-fatality rate (IFR), vaccine efficacy against infection ($VE_{inf}$), and vaccine efficacy against death given infection ($VE_{death}$).

## 6 Discussion

Recent empirical studies have estimated COVID-19 outcomes directly averted by vaccine rollout programs among vaccinated individuals. Here, we define a causal estimand for quantifying directly averted outcomes for vaccination at multiple time points and develop an unbiased estimator. We also examine a popular estimator used by recent empirical studies—the hazard difference estimator (as we call it)—and showed that it is biased relative to the causal estimand when vaccination has a non-null effect, as it fails to incorporate the preferential depletion of susceptibles among the unvaccinated individuals. The simulations performed here, albeit limited, suggest that the bias is substantial for averted infections, as susceptibles were preferentially depleted among the unvaccinated group quickly (due to high effective contacts and high vaccine efficacy against infection). On the other hand, the bias for averted deaths is small when IFR is ≤10% (since survival is similar between vaccinated and unvaccinated individuals), as is the case for many important infections.

Empirical studies frequently used the hazard difference estimator. As a measure, the hazard among (un)vaccinated individuals is restricted to those who have not experienced the outcome between baseline and the start of interval $k$ for $k \geq 1$. Consequently, the interval-specific hazard is subject to differential depletion of susceptibles among the unvaccinated group over time (assuming vaccines have protective effects).[20] However, the hazard difference estimator multiplies the number of vaccinated survivors by the hazard difference between vaccinated and unvaccinated individuals—implicitly assuming that the unvaccinated individuals have the same counterfactual survival as the vaccinated individuals, thereby failing to account for such differential depletion of susceptibles among the unvaccinated group. As our simulations show, the hazard difference estimator overestimates the number of infections averted when susceptibles are preferentially and quickly depleted among unvaccinated individuals given a high reproduction number ($R_0 \approx 3.57$) and a high $VE_{inf}$. The bias is less pronounced for averted deaths when IFR is low, as survival between vaccinated and unvaccinated individuals is similar. Therefore, for COVID-19 studies using the hazard difference estimator, the averted infection estimate is likely more biased than the averted death estimate.

Researchers can use data by aggregated by vaccination status to estimate averted outcomes by following the procedures outlined in **eAppendices 7 and 13**. Most publicly available data from vaccine registries are aggregated by vaccination status due to privacy or other considerations. In these situations, we recommend clearly specifying the intervals to which survival and hazard pertain. An example dataset is shown in **Table S11**, which organizes the hazard and the number of survivors within the same row, specifying the hazard to be one interval ahead of the number of survivors. This arrangement facilitates applying formulas for the alternative unbiased estimator or hazard difference estimator. The table also includes an explanatory footnote clarifying the intervals corresponding to the hazard and the number of survivors.

In the main text, we focus on vaccine-averted outcomes. Some other empirical studies estimated vaccine-avertible deaths—deaths that could have been averted by vaccination, but were not because of a failure to vaccinate. They used the hazard difference estimator for outcomes directly *avertible* by vaccination by multiplying the hazard difference with the number of unvaccinated survivors and summing across weeks.[11,12] **eAppendix 14** defines the estimand, the unbiased estimator, and the hazard difference estimator for outcomes directly avertible, and shows simulation results under the same Scenarios as outlined in **Table 1**. Compared to the causal estimand, the hazard difference estimator could recover a similar value for avertible deaths when IFR is modest (≤10%) or when vaccines are highly effective at preventing death given infection ($VE_{death}$ = 90%) (**Figures S6 & S7**). Note that when estimating directly avertible outcomes (or when comparing to any other counterfactual vaccination strategy than no vaccination), researchers need data disaggregated by vaccination time to use the unbiased estimator.

One major limitation is that the causal estimand and estimators proposed here are developed in the context of an ideal RCT assuming no confounding, no selection bias, perfect compliance, and no other sources of biases. In **eAppendix 13**, we discussed identifying the causal estimand for averted outcomes using observational data aggregated by vaccination status in the absence of confounding. In most observational settings, however, there may be strong confounding by vaccine uptake (i.e., individuals who choose to be vaccinated are also more likely to avoid infection, or elderly individuals are more likely to be vaccinated and also more likely to die from infection). In addition, individuals may reduce protective behaviors after vaccination (i.e., risk compensation),[23] and those who have been infected are also less likely to receive vaccination.[22] Confounding is also a concern in existing studies that estimate averted deaths from observational data, as they adjust for it by stratifying on only simplistic covariates (**Table S1**). Although observational studies violate the assumptions of an ideal RCT, addressing all these violations is beyond the scope of this paper. Future research could use observational data to emulate a target trial similar to the one described here for estimating vaccine-averted outcomes.

In conclusion, motivated by recent empirical studies estimating outcomes directly averted by vaccine rollout programs, we define a causal estimand for directly averted outcomes under interference, which is a lower bound on the total outcomes averted in the entire population when indirect effect is non-negative. We also develop an unbiased estimator in the context of a one-stage RCT and examine the bias of a popular estimator (the hazard difference estimator). The hazard difference estimator is biased relative to the causal estimand when vaccination has a non-null effect because it does not incorporate differential depletion of susceptibles among unvaccinated individuals. Our simulations, albeit limited, show that the hazard difference estimator could substantially overestimate the averted infections when the basic reproduction number and vaccine efficacy against infection are high, while the overestimation is small for averted deaths under modest IFR, which is the case for many important infections.

**eAppendix of "Defining and Estimating Outcomes Directly Averted by a Vaccination Program when Rollout Occurs Over Time"**

**eAppendix 1. Literature review of studies estimating the number of COVID-19 outcomes directly averted among vaccinated individuals or the total number of outcomes averted among both vaccinated and unvaccinated individuals**

Recent studies estimated COVID-19 outcomes averted by vaccine rollout programs either by targeting the outcomes directly averted among vaccinated individuals[1–5] or by targeting the total outcomes averted among both vaccinated and unvaccinated individuals.[6–10] To better understand the estimation methodologies, we reviewed selected literature listed in **Table S1**.

**TABLE S1 |** Summary of selected studies that estimated COVID-19 outcomes averted by vaccine rollout programs

| Year | Authors | Country | Title | COVID-19 Outcomes | Estimand [a] | Estimation methods |
|---|---|---|---|---|---|---|
| 2021 | Haas et al [1] | Israel | Infections, hospitalizations, and deaths averted via a nationwide vaccination campaign using the Pfizer–BioNTech BNT162b2 mRNA COVID-19 vaccine in Israel: a retrospective surveillance study | Infections, hospitalizations, severe or critical hospitalizations, deaths | Direct impact | 1. For each age group and each day, calculate the number of vaccinated individuals at risk of developing the outcome by multiplying the number at risk with the proportion vaccinated. Also calculate the number of unvaccinated individuals by subtracting the numbers of vaccinated individuals from the total population at risk.<br>2. Calculate the hazard rate difference between unvaccinated and vaccinated individuals.<br>3. Estimate the daily averted outcomes by multiplying the number of vaccinated individuals at risk with the hazard rate difference.<br>4. Sum the daily averted outcomes across days within each age group, and then sum across age groups. |
| 2024 | Brault et al [2] | Chile | Direct impact of COVID-19 vaccination in Chile: averted cases, hospitalizations, ICU admissions, and deaths | Cases, hospitalizations, ICU admissions, deaths | Direct impact | 1. For each age group and each week, calculate the number of vaccinated individuals at risk by multiplying the number of individuals at risk with the proportion vaccinated.<br>2. Calculate the incidence rate difference[b] between unvaccinated and vaccinated.<br>3. Estimate the weekly averted outcomes by multiplying the number of vaccinated individuals at risk with the incidence rate difference.<br>4. Sum the weekly averted outcomes across weeks and age groups. |
| 2023 | Santos et al [3] | Brazil | Estimated COVID-19 severe cases and deaths averted in the first year of the vaccination campaign in Brazil: a retrospective observational study | Severe cases and deaths | Direct impact | Same as Haas et al. |
| 2022 | Kayano et al [4] | Japan | Number of averted COVID-19 cases and deaths attributable to reduced risk in vaccinated individuals in Japan | Cases and deaths | Direct impact | Similar to Brault et al, except that calculations in Steps 1 to 3 are stratified by age group, day, and sex. |
| 2023 | Kayano et al [5] | Japan (Tokyo) | Assessing the COVID-19 vaccination program during the Omicron variant (B.1.1.529) epidemic in early 2022, Tokyo | Infections | Direct impact, overall impact[c] | **Direct impact**: Similar to Brault et al<br>**Overall impact**:<br>1. Develop a transmission model based on a renewal process and calibrate the model to the cumulative incidence<br>2. Simulate the counterfactual trajectory under no vaccination by eliminating the vaccination from the fitted transmission model<br>3. Estimate the averted outcomes by taking the difference between observed outcomes and that under the counterfactual scenario of no vaccination |
| 2021 | Vilches et al [6] | USA | Estimating COVID-19 infections, hospitalizations, and deaths following the US vaccination campaigns during the pandemic | Infections, hospitalizations, and deaths | Overall impact | 1. Calibrate an agent-based model to incidence data<br>2. Simulate the trajectory under scenarios with and without vaccination<br>3. Estimate the averted outcomes by taking the difference in outcomes between scenarios with and without vaccination |
| 2022 | Schneider et al [7] | USA | Impact of U.S. COVID-19 Vaccination Efforts: An Update on Averted Deaths, Hospitalizations, and Health | Infections, hospitalizations, and deaths | Overall impact | Similar to Vilches et al. |

| | | | Care Costs Through March 2022 | | | |
|---|---|---|---|---|---|---|
| 2022 | Gavish et al [8] | Israel | Population-level implications of the Israeli booster campaign to curtail COVID-19 resurgence | Cases, severe cases, and deaths | Overall impact[d] | Similar to Vilches et al., except that they considered scenarios with or without booster to investigate the outcomes averted by booster. |
| 2021 | Shoukat et al [10] | USA (New York City) | Lives saved and hospitalizations averted by COVID-19 | Cases, hospitalizations, and deaths | Overall impact | 1. Calibrate an agent-based model to incidence data<br>2. Simulate the trajectory under counterfactual scenario without vaccination<br>3. Estimate the averted outcomes by taking the difference in outcomes between observed outcomes and that under the counterfactual scenario of no vaccination |
| 2022 | Watson et al [9] | 185 countries | Global impact of the first year of COVID-19 vaccination: a mathematical modelling study | Deaths[e] | Overall impact[f] | Similar to Vilches et al., except that they used an age-structured compartmental model |

[a] Direct impact refers to the number of outcomes directly averted among vaccinated individuals; overall impact refers to the total number of outcomes averted among both vaccinated and unvaccinated individuals.

[b] Incidence rate among vaccinated individuals was implicitly conditional on not having developed the outcome until time of vaccination and thus was a hazard measure.

[c] The overall impact corresponds to what the authors referred to as the "total effect" in their paper, described as "the total effect at the population level, consisting of the vaccine-induced protection that is conferred directly and indirectly."

[d] The authors also estimated the indirect effect of boosters.

[e] The authors also estimated all-cause mortality averted.

[f] The authors also estimated the number of outcomes averted via indirect effect.

**eAppendix 2. Defining individual, group, and population average potential outcomes**

**1 Mixed individual assignment strategy**

Let $\boldsymbol{\rho} = \{q+1, d; \rho_0, \ldots, \rho_q\}$, where $\sum_{x=0}^{q+1} \rho_x = 1$, be a parameterization that governs the distribution of vaccination times $\mathbf{X}$. Let $K_x \equiv N \cdot \rho_x \equiv \sum_j I(X_j = x)$ for $x \in \{0, \ldots, q+1\}$ and $j \in \{1, \ldots, N\}$. Define $\boldsymbol{\rho}$ to be a *mixed individual assignment strategy* if $K_x$ is fixed under $\boldsymbol{\rho}$, with $0 < K_x < N$ for all $x \in \{0, \ldots, q+1\}$ and each of the $\frac{N!}{\prod_{x=0}^{q+1}(K_x!)}$ possible individual assignments receiving equal probability.[11,12]

**2 Type A parameterization with categorical treatment variable**

Hudgens and Halloran[11] considered a mixed assignment strategy[12] for binary treatment variable (i.e., vaccination or non-vaccination) in defining the individual average potential outcome; their approach was referred to as Type A parameterization by VanderWeele and Tchetgen Tchetgen.[13] Here, we define the individual average potential outcomes using Type A parameterization for categorical treatment variable.

Assume $\boldsymbol{\rho}$ is a mixed assignment strategy. Let $\pi(\mathbf{X} = \mathbf{x}; \boldsymbol{\rho})$ denote the probability that the group is assigned with $\mathbf{X}$ given parameter $\boldsymbol{\rho}$. The vaccination times $\mathbf{X}$ are randomly assigned conditional on $\{K_0, \ldots, K_{q+1}\}$ with probability mass function:

$$\pi(\mathbf{X} = \mathbf{x}; \boldsymbol{\rho}) = \frac{\prod_{x=0}^{q+1} I\left(\sum_{j=1}^{N} I\left(X_j = x\right) = K_x\right)}{\frac{N!}{\prod_{x=0}^{q+1}(K_x!)}}$$

where $\frac{N!}{\prod_{x=0}^{q+1}(K_x!)}$ is the number of ways to assign exactly $K_0, \ldots, K_{q+1}$ individuals to $x = 0, \ldots, q+1$, respectively.

**3 Individual and group average potential outcomes**

By the beginning of interval $q+1$, the *individual average potential outcome*[11] for individual $j$ with time of vaccination $x$ in the group under strategy $\boldsymbol{\rho}$ is:

$$\bar{Y}_{q+1,j}(x;\boldsymbol{\rho}) \equiv \sum_{\boldsymbol{\omega}\in\mathcal{X}(N-1)} Y_{q+1,j}\left(\mathbf{x}_{-j}=\boldsymbol{\omega}, x_j = x\right) \Pr_{\boldsymbol{\rho}}\left(\mathbf{X}_{-j}=\boldsymbol{\omega} | X_j = x\right)$$

where $\Pr_{\boldsymbol{\rho}}\left(\mathbf{X}_{-j}=\boldsymbol{\omega}|X_j=x\right) = \frac{\pi\left(\mathbf{X}_{-j}=\boldsymbol{\omega}, X_j=x;\boldsymbol{\rho}\right)}{\sum_{\boldsymbol{\omega}'\in\mathcal{X}(N-1)}\pi\left(\mathbf{X}_{-j}=\boldsymbol{\omega}', X_j=x;\boldsymbol{\rho}\right)}$. See Definition 2 in VanderWeele and Tchetgen Tchetgen for the analog with a binary treatment variable.[13] As they noted,[13] if $\bar{Y}_{q+1,j}$ is defined under Type A parameterization, the proportion of other individuals assigned to, for example, $x=0$, varies depending on whether individual $j$ is assigned to $x=0$. In other words, the proportion $\rho_0$ for others is not held fixed in $\bar{Y}_{q+1,j}(0;\boldsymbol{\rho})$ compared to $\bar{Y}_{q+1,j}(k;\boldsymbol{\rho})$ for some $k>0$, so that the so-called direct effect (which can be written as $DE_{q+1}((k,0);\boldsymbol{\rho}) = \bar{Y}_{q+1}(k;\boldsymbol{\rho}) - \bar{Y}_{q+1}(0;\boldsymbol{\rho}) = \frac{\sum_{j=1}^{N}\left(\bar{Y}_{q+1,j}(k;\boldsymbol{\rho}) - \bar{Y}_{q+1,j}(0;\boldsymbol{\rho})\right)}{N}$)[11] does not merit the label "direct effect." However, this issue is negligible under a large group size $N$.

The group average potential outcome is:

$$\bar{Y}_{q+1}(x;\boldsymbol{\rho}) = \frac{\sum_{j=1}^{N} \bar{Y}_{q+1,j}\left(x_j = x;\boldsymbol{\rho}\right)}{N}.$$

**eAppendix 3. Causal estimand for outcomes directly averted by vaccination**

**1 Extending the causal estimand to an arbitrary number of vaccination times**

For $q \in \mathbb{N}$, the causal estimand for outcomes directly averted by $\boldsymbol{\rho} = \{q+1, d; \rho_0, \ldots, \rho_q\}$ where $\sum_{x=0}^{q+1} \rho_x = 1$, compared to no vaccination $\boldsymbol{\phi} = \{q+1, d; \mathbf{0}\}$ is:

*Definition S1 (Causal estimand for directly averted outcomes)*

$$\delta_{q+1}^{D}(\boldsymbol{\phi}, \boldsymbol{\rho}) = N \cdot \sum_{k=0}^{q} \rho_k \cdot \left[\bar{Y}_{q+1}(q+1; \boldsymbol{\rho}) - \bar{Y}_{q+1}(k; \boldsymbol{\rho})\right] \quad \text{(S1)}$$

$$= N \cdot \sum_{k=0}^{q} \rho_k \cdot DE_{q+1}\big((q+1, k); \boldsymbol{\rho}\big)$$

$\delta_{q+1}^{D}(\boldsymbol{\phi}, \boldsymbol{\rho})$ in equation (S1) can be re-written as:

$$\delta_{q+1}^{D}(\boldsymbol{\phi}, \boldsymbol{\rho})$$

$$= N \cdot \sum_{k=0}^{q} \rho_k \cdot \left[\sum_{l=1}^{q+1} \Delta\bar{Y}_l(q+1; \boldsymbol{\rho}) - \sum_{l=1}^{q+1} \Delta\bar{Y}_l(k; \boldsymbol{\rho})\right]$$

$$= N \cdot \Bigg[\rho_0 \cdot \left(\sum_{l=1}^{q+1} \Delta\bar{Y}_l(q+1; \boldsymbol{\rho}) - \sum_{l=1}^{q+1} \Delta\bar{Y}_l(0; \boldsymbol{\rho})\right) + \rho_1 \cdot \left(\sum_{l=1}^{q+1} \Delta\bar{Y}_l(q+1; \boldsymbol{\rho}) - \sum_{l=1}^{q+1} \Delta\bar{Y}_l(1; \boldsymbol{\rho})\right) + \rho_2$$
$$\cdot \left(\sum_{l=1}^{q+1} \Delta\bar{Y}_l(q+1; \boldsymbol{\rho}) - \sum_{l=1}^{q+1} \Delta\bar{Y}_l(2; \boldsymbol{\rho})\right) + \cdots + \rho_h \cdot \left(\sum_{l=1}^{q+1} \Delta\bar{Y}_l(q+1; \boldsymbol{\rho}) - \sum_{l=1}^{q+1} \Delta\bar{Y}_l(h; \boldsymbol{\rho})\right) + \cdots + \rho_q$$
$$\cdot \left(\sum_{l=1}^{q+1} \Delta\bar{Y}_l(q+1; \boldsymbol{\rho}) - \sum_{l=1}^{q+1} \Delta\bar{Y}_l(q; \boldsymbol{\rho})\right)\Bigg]$$

$$= N \cdot \Bigg[\rho_0 \cdot \left(\sum_{l=1}^{q+1} \Delta\bar{Y}_l(q+1; \boldsymbol{\rho}) - \sum_{l=1}^{q+1} \Delta\bar{Y}_l(0; \boldsymbol{\rho})\right) + \rho_1 \cdot \left(\sum_{l=2}^{q+1} \Delta\bar{Y}_l(q+1; \boldsymbol{\rho}) - \sum_{l=2}^{q+1} \Delta\bar{Y}_l(1; \boldsymbol{\rho})\right) + \rho_2$$
$$\cdot \left(\sum_{l=3}^{q+1} \Delta\bar{Y}_l(q+1; \boldsymbol{\rho}) - \sum_{l=3}^{q+1} \Delta\bar{Y}_l(2; \boldsymbol{\rho})\right) + \cdots + \rho_{h-1} \cdot \left(\sum_{l=h}^{q+1} \Delta\bar{Y}_l(q+1; \boldsymbol{\rho}) - \sum_{l=h}^{q+1} \Delta\bar{Y}_l(h-1; \boldsymbol{\rho})\right) + \cdots + \rho_q$$
$$\cdot \left(\Delta\bar{Y}_{q+1}(q+1; \boldsymbol{\rho}) - \Delta\bar{Y}_{q+1}(q; \boldsymbol{\rho})\right)\Bigg]$$

$$= N \cdot \Bigg[\rho_0 \cdot \big(\Delta\bar{Y}_1(q+1;\boldsymbol{\rho}) - \Delta\bar{Y}_1(0;\boldsymbol{\rho})\big) + (\rho_0+\rho_1) \cdot \left(\Delta\bar{Y}_2(q+1;\boldsymbol{\rho}) - \frac{\rho_0\Delta\bar{Y}_2(0;\boldsymbol{\rho}) + \rho_1\Delta\bar{Y}_2(1;\boldsymbol{\rho})}{\rho_0+\rho_1}\right) + (\rho_0+\rho_1+\rho_2)$$

$$\cdot \left(\Delta\bar{Y}_3(q+1;\boldsymbol{\rho}) - \frac{\rho_0\Delta\bar{Y}_3(0;\boldsymbol{\rho}) + \rho_1\Delta\bar{Y}_3(1;\boldsymbol{\rho}) + \rho_2\Delta\bar{Y}_3(2;\boldsymbol{\rho})}{\rho_0+\rho_1+\rho_2}\right) + \cdots + \left(\sum_{k=0}^{h-1}\rho_k\right)$$

$$\cdot \left(\Delta\bar{Y}_h(q+1;\boldsymbol{\rho}) - \frac{\sum_{k=0}^{h-1}\rho_k\Delta\bar{Y}_h(k;\boldsymbol{\rho})}{\sum_{k=0}^{h-1}\rho_k}\right) + \cdots + \left(\sum_{k=0}^{q}\rho_k\right) \cdot \left(\Delta\bar{Y}_{q+1}(q+1;\boldsymbol{\rho}) - \frac{\sum_{k=0}^{q}\rho_k\Delta\bar{Y}_{q+1}(k;\boldsymbol{\rho})}{\sum_{k=0}^{q}\rho_k}\right)\Bigg]$$

$$= N \cdot \sum_{l=1}^{q+1}\left[\left(\sum_{k=0}^{l-1}\rho_k\right) \cdot \left(\Delta\bar{Y}_l(q+1;\boldsymbol{\rho}) - \frac{\sum_{k=0}^{l-1}\rho_k\Delta\bar{Y}_l(k;\boldsymbol{\rho})}{\sum_{k=0}^{l-1}\rho_k}\right)\right]. \qquad \text{(S2)}$$

We assume individuals with $x = l < q + 1$ have had the same historical probability of developing the outcome as those with $x = q + 1$, up until $l$. Therefore, from the third to the forth line, $\Delta\bar{Y}_1(q+1;\boldsymbol{\rho})$ cancels out $\Delta\bar{Y}_1(1;\boldsymbol{\rho})$, since $x = 1$ is unvaccinated before the beginning of interval 1; similarly, $\sum_{l=1}^{2}\Delta\bar{Y}_l(q+1;\boldsymbol{\rho})$ cancels out $\sum_{l=1}^{2}\Delta\bar{Y}_l(2;\boldsymbol{\rho})$, since $x = 2$ is unvaccinated before the beginning of interval 2, and so on for other vaccination groups. In words, equation (S2) is the proportion ever-vaccinated (i.e., $\sum_{k=0}^{l-1}\rho_k$) multiplied with the difference between the period incidence $\Delta\bar{Y}_l(\cdot)$ among individuals assigned to no vaccination and the average period incidence among individuals ever-vaccinated, summed across time intervals $l \in (1, \dots, q+1)$.

Furthermore, as we assume individuals with $x = l < q + 1$ have had the same historical probability of developing outcome as those with $x = q + 1$, up until $l$, we have $\Delta\bar{Y}_l(q+1;\boldsymbol{\rho}) = \frac{\sum_{k=l}^{q+1}\rho_k\Delta\bar{Y}_l(k;\boldsymbol{\rho})}{\sum_{k=l}^{q+1}\rho_k}$, such that equation (S2) can be written as:

$$\delta_{q+1}^{D}(\boldsymbol{\phi},\boldsymbol{\rho}) = N \cdot \sum_{l=1}^{q+1}\left[\left(\sum_{k=0}^{l-1}\rho_k\right) \cdot \left(\frac{\sum_{k=l}^{q+1}\rho_k\Delta\bar{Y}_l(k;\boldsymbol{\rho})}{\sum_{k=l}^{q+1}\rho_k} - \frac{\sum_{k=0}^{l-1}\rho_k\Delta\bar{Y}_l(k;\boldsymbol{\rho})}{\sum_{k=0}^{l-1}\rho_k}\right)\right] \qquad \text{(S3)}$$

$$= N \cdot \left[\sum_{l=1}^{q+1}\left(\sum_{k=0}^{l-1}\rho_k\right) \cdot \frac{\sum_{k=l}^{q+1}\rho_k\Delta\bar{Y}_l(k;\boldsymbol{\rho})}{\sum_{k=l}^{q+1}\rho_k} - \sum_{l=1}^{q+1}\sum_{k=0}^{l-1}\rho_k\Delta\bar{Y}_l(k;\boldsymbol{\rho})\right] \qquad \text{(S4)}$$

Writing $\Delta\bar{Y}_l(q+1;\boldsymbol{\rho}) = \frac{\sum_{k=l}^{q+1}\rho_k\Delta\bar{Y}_l(k;\boldsymbol{\rho})}{\sum_{k=l}^{q+1}\rho_k}$ also has the advantage of preserving data because they are equivalent in an ideal trial while $\frac{\sum_{k=l}^{q+1}\rho_k\Delta\bar{Y}_l(k;\boldsymbol{\rho})}{\sum_{k=l}^{q+1}\rho_k}$ can be estimated with a larger sample.

## 2 An example: vaccination at two time points

To be explicit, consider vaccination at two time points. Recall equation (3) that defines outcomes directly averted by $\boldsymbol{\rho} = \{2, d; \rho_0, \rho_1\}$, compared to no vaccination $\boldsymbol{\phi} = \{2, d; \mathbf{0}\}$:

$$\delta_2^D(\boldsymbol{\phi}, \boldsymbol{\rho})$$

$$= N \cdot [\rho_0 \cdot (\bar{Y}_2(2;\boldsymbol{\rho}) - \bar{Y}_2(0;\boldsymbol{\rho})) + \rho_1 \cdot (\bar{Y}_2(2;\boldsymbol{\rho}) - \bar{Y}_2(1;\boldsymbol{\rho}))$$

$$= N \cdot \{\rho_0 \cdot [(\Delta\bar{Y}_1(2;\boldsymbol{\rho}) + \Delta\bar{Y}_2(2;\boldsymbol{\rho})) - (\Delta\bar{Y}_1(0;\boldsymbol{\rho}) + \Delta\bar{Y}_2(0;\boldsymbol{\rho}))] + \rho_1 \cdot [(\Delta\bar{Y}_1(2;\boldsymbol{\rho}) + \Delta\bar{Y}_2(2;\boldsymbol{\rho})) - (\Delta\bar{Y}_1(1;\boldsymbol{\rho}) + \Delta\bar{Y}_2(1;\boldsymbol{\rho}))]\}$$

$$= N \cdot \{\rho_0 \cdot [(\Delta\bar{Y}_1(2;\boldsymbol{\rho}) + \Delta\bar{Y}_2(2;\boldsymbol{\rho})) - (\Delta\bar{Y}_1(0;\boldsymbol{\rho}) + \Delta\bar{Y}_2(0;\boldsymbol{\rho}))] + \rho_1 \cdot (\Delta\bar{Y}_2(2;\boldsymbol{\rho}) - \Delta\bar{Y}_2(1;\boldsymbol{\rho}))\}$$

$$= N \cdot \left\{\rho_0 \cdot (\Delta\bar{Y}_1(2;\boldsymbol{\rho}) - \Delta\bar{Y}_1(0;\boldsymbol{\rho})) + (\rho_0 + \rho_1)\left[\Delta\bar{Y}_2(2;\boldsymbol{\rho}) - \left(\frac{\rho_0\Delta\bar{Y}_2(0;\boldsymbol{\rho}) + \rho_1\Delta\bar{Y}_2(1;\boldsymbol{\rho})}{\rho_0 + \rho_1}\right)\right]\right\}$$

Assume $\Delta\bar{Y}_1(2;\boldsymbol{\rho}) = \dfrac{\rho_1\Delta\bar{Y}_1(1;\boldsymbol{\rho}) + \rho_2\Delta\bar{Y}_1(2;\boldsymbol{\rho})}{\rho_1 + \rho_2} \longrightarrow$

$$= N \cdot \left\{\rho_0 \cdot \left(\frac{\rho_1\Delta\bar{Y}_1(1;\boldsymbol{\rho}) + \rho_2\Delta\bar{Y}_1(2;\boldsymbol{\rho})}{\rho_1 + \rho_2} - \Delta\bar{Y}_1(0;\boldsymbol{\rho})\right) + [(\rho_0 + \rho_1) \cdot \Delta\bar{Y}_2(2;\boldsymbol{\rho}) - (\rho_0\Delta\bar{Y}_2(0;\boldsymbol{\rho}) + \rho_1\Delta\bar{Y}_2(1;\boldsymbol{\rho}))]\right\}$$

$$= N \cdot \sum_{l=1}^{2}\left[\left(\sum_{k=0}^{l-1}\rho_k\right) \cdot \left(\frac{\sum_{k=l}^{2}\rho_k\Delta\bar{Y}_l(k;\boldsymbol{\rho})}{\sum_{k=l}^{2}\rho_k} - \frac{\sum_{k=0}^{l-1}\rho_k\Delta\bar{Y}_l(k;\boldsymbol{\rho})}{\sum_{k=0}^{l-1}\rho_k}\right)\right]$$

$$= N \cdot \left[\sum_{l=1}^{2}\left(\sum_{k=0}^{l-1}\rho_k\right) \cdot \left(\frac{\sum_{k=l}^{2}\rho_k\Delta\bar{Y}_l(k;\boldsymbol{\rho})}{\sum_{k=l}^{2}\rho_k}\right) - \sum_{l=1}^{2}\sum_{k=0}^{l-1}\rho_k\Delta\bar{Y}_l(k;\boldsymbol{\rho})\right]$$

**eAppendix 4 Conditions under which directly averted outcomes is a lower bound on overall averted outcomes**

To identify conditions under which the number of directly averted outcomes among vaccinated individuals is a lower bound on the total number of outcomes averted, we first define the "overall impact" estimand to quantify outcomes averted in both vaccinated and unvaccinated individuals. Previously, we defined the overall impact estimand for vaccination at a single time point,[14] for which we now extend to multiple time points. First, we need to define the indirect, total, and overall *effects* for vaccination at multiple time points.

**1 Indirect, total, and overall effects for vaccination at multiple time points**

When vaccination occurs at multiple time points, there could be separate versions of indirect and total effects under each possible value $x \in \{0, \dots, q+1\}$.

**1.1 Indirect effect**

In general, indirect effect can be a contrast between two strategies $\boldsymbol{\rho}$ and $\widetilde{\boldsymbol{\rho}}$, conditional on $x$. In notation, the indirect effect comparing probability of having developed the outcome by beginning of interval $q+1$ for an individual assigned to vaccination time $x$ when the group follows strategy $\boldsymbol{\rho}$ versus $\widetilde{\boldsymbol{\rho}}$ is:

$$IE_{q+1}\big(x; (\boldsymbol{\rho}, \widetilde{\boldsymbol{\rho}})\big) = \bar{Y}_{q+1}(x; \boldsymbol{\rho}) - \bar{Y}_{q+1}(x; \widetilde{\boldsymbol{\rho}}). \quad \text{(S5)}$$

**1.2 Total effect**

Total effect can be any contrast between any combinations of $(x, \boldsymbol{\rho})$ and $(x', \widetilde{\boldsymbol{\rho}})$. In notation, the total effect comparing probability of having developed the outcome by beginning of $q+1$ for an individual assigned to vaccination time $x$ when the group follows strategy $\boldsymbol{\rho}$ versus an individual with $x'$ when the group follows strategy $\widetilde{\boldsymbol{\rho}}$ is:

$$TE_{q+1}\big((x, x'); (\boldsymbol{\rho}, \widetilde{\boldsymbol{\rho}})\big) = \bar{Y}_{q+1}(x; \boldsymbol{\rho}) - \bar{Y}_{q+1}(x'; \widetilde{\boldsymbol{\rho}}). \quad \text{(S6)}$$

By definition, total effect is the sum of direct and indirect effects:

$TE_{q+1}\big((x,x');(\boldsymbol{\rho},\widetilde{\boldsymbol{\rho}})\big) = \bar{Y}_{q+1}(x;\boldsymbol{\rho}) - \bar{Y}_{q+1}(x';\widetilde{\boldsymbol{\rho}})$

$= \bar{Y}_{q+1}(x;\boldsymbol{\rho}) - \bar{Y}_{q+1}(x';\boldsymbol{\rho}) + \bar{Y}_{q+1}(x';\boldsymbol{\rho}) - \bar{Y}_{q+1}(x';\widetilde{\boldsymbol{\rho}})$

$= DE_{q+1}((x,x');\boldsymbol{\rho}) + IE_{q+1}\big(x';(\boldsymbol{\rho},\widetilde{\boldsymbol{\rho}})\big).$ (S7)

Hudgens and Halloran[11] also partitioned the total effect into the sum of direct and indirect effects for vaccination at a single time point.

**1.3 Overall effect**

The overall effect comparing probability of having developed the outcome by beginning of $q+1$ for a typical individual in the group following strategy $\boldsymbol{\rho}$ versus $\widetilde{\boldsymbol{\rho}}$ is:

$$OE_{q+1}(\boldsymbol{\rho},\widetilde{\boldsymbol{\rho}}) = \bar{Y}_{q+1}(\boldsymbol{\rho}) - \bar{Y}_{q+1}(\widetilde{\boldsymbol{\rho}}). \quad \text{(S8)}$$

**2 Partitioning overall effect when comparing no vaccination to some vaccination**

Consider the comparison between no vaccination and a vaccination strategy $\boldsymbol{\rho}$, as our goal is to quantify the overall impact of the strategy. Previously, Sobel[12] and Hudgens and Halloran[11] noted that when comparing no vaccination to vaccination at baseline, overall effect is the weighted sum of total and indirect effects. Here, we extend the partitioning of overall effect to vaccination at multiple time points (Theorem S1).

*Theorem S1 (Overall effect partitioning for no vaccination vs.* $\boldsymbol{\rho}$*)* Let $\boldsymbol{\phi} = \{q+1, d; \mathbf{0}\}$ for $q \in \mathbb{N}$ and $d \in \mathbb{Z}^{+}$. Then

$$OE_{q+1}(\boldsymbol{\phi},\boldsymbol{\rho}) = \left[\sum_{k=0}^{q} \rho_k \cdot TE_{q+1}\big((q+1,k);(\boldsymbol{\phi},\boldsymbol{\rho})\big)\right] + \rho_{q+1} \cdot IE_{q+1}\big(q+1;(\boldsymbol{\phi},\boldsymbol{\rho})\big).$$

*Proof.* For $q \in \mathbb{N}$, we have

$OE_{q+1}(\boldsymbol{\phi},\boldsymbol{\rho}) = \bar{Y}_{q+1}(\boldsymbol{\phi}) - \bar{Y}_{q+1}(\boldsymbol{\rho})$

All is unvaccinated under $\boldsymbol{\phi} \longrightarrow$

$$= \bar{Y}_{q+1}(q+1;\boldsymbol{\phi}) - \bar{Y}_{q+1}(\boldsymbol{\rho})$$

By definition, $\sum_{k=0}^{q+1} \rho_k = 1 \longrightarrow$

$$= \left[\sum_{k=0}^{q} \rho_k \cdot \bar{Y}_{q+1}(q+1;\boldsymbol{\phi}) + \rho_{q+1} \cdot \bar{Y}_{q+1}(q+1;\boldsymbol{\phi})\right]$$

$$- \left[\sum_{k=0}^{q} \rho_k \cdot \bar{Y}_{q+1}(k;\boldsymbol{\rho}) + \rho_{q+1} \cdot \bar{Y}_{q+1}(q+1;\boldsymbol{\rho})\right]$$

$$= \sum_{k=0}^{q} \rho_k \cdot \left(\bar{Y}_{q+1}(q+1;\boldsymbol{\phi}) - \bar{Y}_{q+1}(k;\boldsymbol{\rho})\right) + \rho_{q+1} \cdot \left(\bar{Y}_{q+1}(q+1;\boldsymbol{\phi}) - \bar{Y}_{q+1}(q+1;\boldsymbol{\rho})\right)$$

$$= \sum_{k=0}^{q} \rho_k \cdot TE_{q+1}\big((q+1,k);(\boldsymbol{\phi},\boldsymbol{\rho})\big) + \rho_{q+1} \cdot IE_{q+1}\big(q+1;(\boldsymbol{\phi},\boldsymbol{\rho})\big)$$

Consider $q+1=2$. We have $OE_2(\boldsymbol{\phi},\boldsymbol{\rho}) = \rho_0 TE_2\big((2,0);(\boldsymbol{\phi},\boldsymbol{\rho})\big) + \rho_1 TE_2\big((2,1);(\boldsymbol{\phi},\boldsymbol{\rho})\big) + \rho_2 IE_2\big(2;(\boldsymbol{\phi},\boldsymbol{\rho})\big)$. The partitioning is graphically illustrated in **Figure S1**.

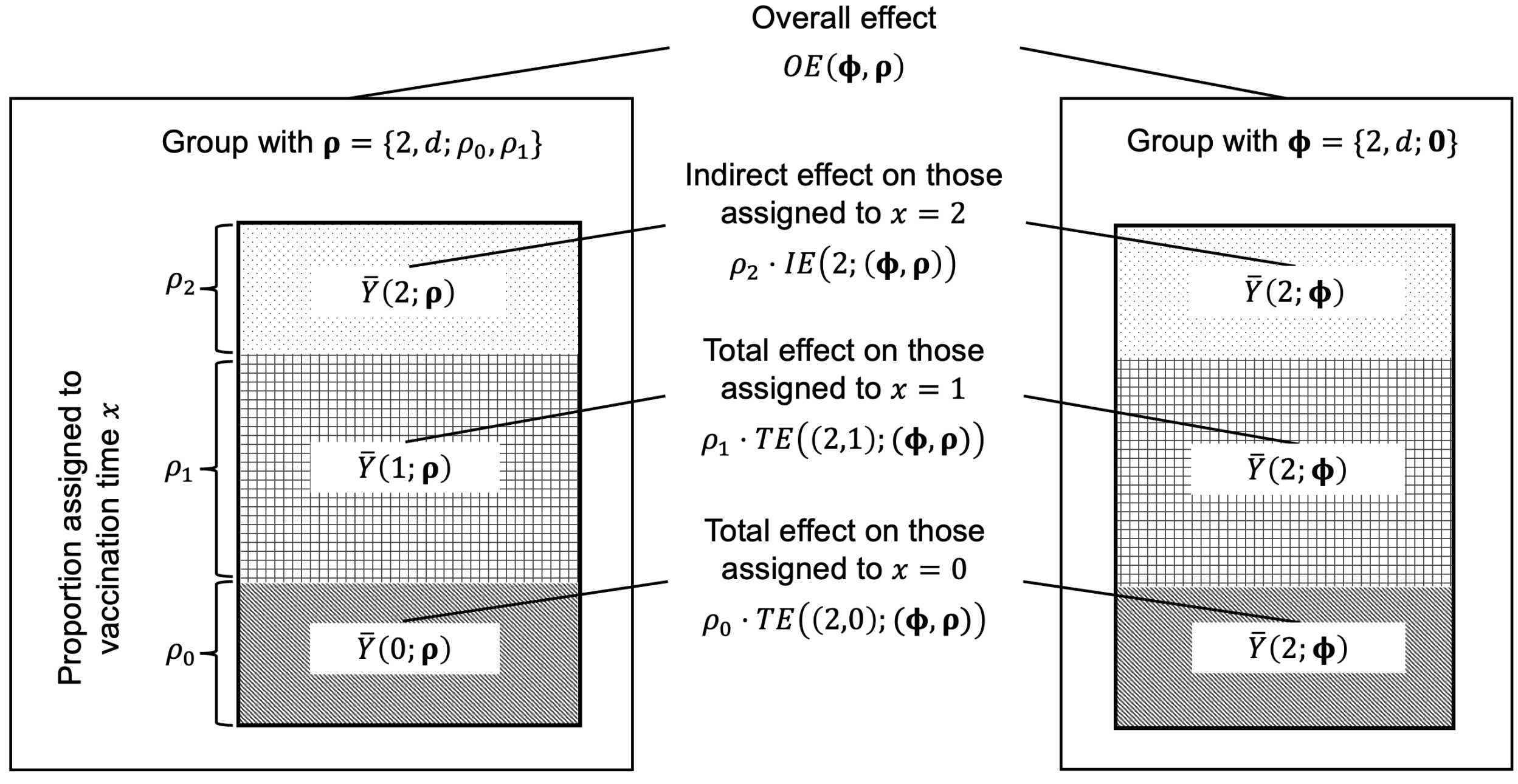

**FIGURE S1 | Graphical illustration on partitioning overall effect.** The two rectangles represent a pair of counterfactuals wherein the group follow strategy $\boldsymbol{\rho} = \{2, d; \rho_0, \rho_1\}$ or $\boldsymbol{\phi} = \{2, d; \mathbf{0}\}$. Individuals fall into three categories based on their assigned vaccination time: 1) The dotted region represents those who are assigned to $x = 2$ and for whom $IE(2; \boldsymbol{\phi}, \boldsymbol{\rho})$ is the difference between the counterfactuals; 2) the gridded region represents those who are assigned to $x = 1$ and for whom $TE\big((2,1); (\boldsymbol{\phi}, \boldsymbol{\rho})\big)$ is the difference between the counterfactuals; and 3) the stripped region represents those who are assigned to $x = 0$ and for whom $TE\big((2,0); (\boldsymbol{\phi}, \boldsymbol{\rho})\big)$ is the difference between the counterfactuals. Theorem S1 shows that $OE(\boldsymbol{\phi}, \boldsymbol{\rho})$ is a weighted average of three effects: 1) $IE\big(2; (\boldsymbol{\phi}, \boldsymbol{\rho})\big)$, 2) $TE\big((2,1); (\boldsymbol{\phi}, \boldsymbol{\rho})\big)$, and 3) $TE\big((2,0); (\boldsymbol{\phi}, \boldsymbol{\rho})\big)$, each weighted by the proportion of individuals for whom the effect is in operation respectively: 1) $\rho_2$ for those assigned to $x = 2$, 2) $\rho_1$ for those assigned to $x = 1$, and 3) $\rho_0$ for those assigned to $x = 0$. The time interval notation suppressed to reduce clutter.

## 3 Defining and partitioning overall impact for vaccination at multiple time points

### 3.1 Overall impact

Now, we use the above definition of overall effect to define the overall impact estimand and apply Theorem S1 to partition overall impact into components of direct and indirect effects.

The total number of outcomes averted among both vaccinated and unvaccinated individuals by strategy $\boldsymbol{\rho} = \{q + 1, d; \rho_0, \dots, \rho_q\}$, compared to no vaccination $\boldsymbol{\phi} = \{q + 1, d; \mathbf{0}\}$, is:

*Definition S2 (Overall impact estimand for quantifying the total number of outcomes averted)*

$$\delta^O_{q+1}(\boldsymbol{\phi}, \boldsymbol{\rho}) = N \cdot OE_{q+1}(\boldsymbol{\phi}, \boldsymbol{\rho}).$$

$\delta^O_{q+1}(\boldsymbol{\phi}, \boldsymbol{\rho})$ addresses the causal question: how many outcomes have been averted among both vaccinated and unvaccinated individuals under strategy $\boldsymbol{\rho}$ compared to no vaccination $\boldsymbol{\phi}$? The mathematical modeling studies[5–10] listed in **Table S1** have targeted $\delta^O_{q+1}(\boldsymbol{\phi}, \boldsymbol{\rho})$ to estimate total outcomes averted by simulating the epidemic trajectory under no vaccination ($\boldsymbol{\phi}$) and comparing it with the trajectory under a particular strategy $\boldsymbol{\rho}$ (or with the observed outcomes).

**3.2 Overall impact partitioning**

By Theorem S1 and the definition from equation (S7) $TE_{q+1}\big((q+1,k);(\boldsymbol{\phi},\boldsymbol{\rho})\big) = IE_{q+1}\big(q+1;(\boldsymbol{\phi},\boldsymbol{\rho})\big) + DE_{q+1}\big((q+1,k);\boldsymbol{\rho}\big)$ for $k \in \{0,\dots,q\}$, we decompose $\delta^{O}_{q+1}(\boldsymbol{\phi},\boldsymbol{\rho})$:

$$\delta^{O}_{q+1}(\boldsymbol{\phi},\boldsymbol{\rho}) = N \cdot OE_{q+1}(\boldsymbol{\phi},\boldsymbol{\rho})$$

$$= N\left[\sum_{k=0}^{q} \rho_k \cdot TE_{q+1}\big((q+1,k);(\boldsymbol{\phi},\boldsymbol{\rho})\big) + \rho_{q+1} \cdot IE_{q+1}\big(q+1;(\boldsymbol{\phi},\boldsymbol{\rho})\big)\right]$$

$$= N\left[\sum_{k=0}^{q} \rho_k \cdot \Big(DE_{q+1}\big((q+1,k);\boldsymbol{\rho}\big) + IE_{q+1}\big(q+1;(\boldsymbol{\phi},\boldsymbol{\rho})\big)\Big) + \rho_{q+1} \cdot IE_{q+1}\big(q+1;(\boldsymbol{\phi},\boldsymbol{\rho})\big)\right]$$

$$= N\left[\left(\sum_{k=0}^{q} \rho_k \cdot DE_{q+1}\big((q+1,k);\boldsymbol{\rho}\big)\right) + IE_{q+1}\big(q+1;(\boldsymbol{\phi},\boldsymbol{\rho})\big)\right] \qquad \text{(S9)}$$

Substituting the definition $\delta^{D}_{q+1}(\boldsymbol{\phi},\boldsymbol{\rho}) = N \cdot \sum_{k=0}^{q} \rho_k \cdot DE_{q+1}\big((q+1,k);\boldsymbol{\rho}\big)$ into the last line of equation (S9), we have:

$$\delta^{O}_{q+1}(\boldsymbol{\phi},\boldsymbol{\rho}) = \delta^{D}_{q+1}(\boldsymbol{\phi},\boldsymbol{\rho}) + N \cdot IE_{q+1}\big(q+1;(\boldsymbol{\phi},\boldsymbol{\rho})\big)$$

Therefore, $\delta^{O}_{q+1}(\boldsymbol{\phi},\boldsymbol{\rho}) \geq \delta^{D}_{q+1}(\boldsymbol{\phi},\boldsymbol{\rho})$ if and only if $IE_{q+1}\big(q+1;(\boldsymbol{\phi},\boldsymbol{\rho})\big) \geq 0$.

**eAppendix 5. Unbiased estimator for the causal estimand in a one-stage randomized controlled trial**

Recall equation (S1): $\delta_{q+1}^{D}(\boldsymbol{\phi}, \boldsymbol{\rho}) = N \cdot \sum_{k=0}^{q} \rho_k \cdot \left[\bar{Y}_{q+1}(q+1; \boldsymbol{\rho}) - \bar{Y}_{q+1}(k; \boldsymbol{\rho})\right]$. To identify $\delta_{q+1}^{D}(\boldsymbol{\phi}, \boldsymbol{\rho})$, consider the following assumptions.

*Assumption 1.* $\boldsymbol{\rho}$ is a mixed assignment strategy.

Recall the definition of mixed assignment strategy in **eAppendix 2**: $\boldsymbol{\rho}$ is a *mixed individual assignment strategy* if $K_x$ is fixed under $\boldsymbol{\rho}$, with $0 < K_x < N$ for all $x \in \{0, \dots, q+1\}$ and each of the $\frac{N!}{\prod_{x=0}^{q+1}(K_x!)}$ possible individual assignments receiving equal probability.[11,12]

*Assumption 2 (consistency).*[26]

$$\text{If } \mathbf{X} = \mathbf{x}, \text{ then } Y_{q+1,j}(\mathbf{x}) = Y_{q+1,j}$$

for all $j$.

To identify $\delta_{q+1}^{D}(\boldsymbol{\phi}, \boldsymbol{\rho})$, we can make use of the following theorem.

*Theorem S2 (Unbiased estimator for the causal estimand in a one-stage RCT).* Let $\boldsymbol{\rho} = \{q+1, d; \rho_0, \dots, \rho_q\}$, where $\sum_{x=0}^{q+1} \rho_x = 1$, and $\boldsymbol{\phi} = \{q+1, d; \mathbf{0}\}$ for $q \in \mathbb{N}$ and $d \in \mathbb{Z}^{+}$. Let

$$\widehat{\delta_{q+1}}^{D}(\boldsymbol{\phi}, \boldsymbol{\rho}) = N \cdot \sum_{k=0}^{q} \rho_k \cdot \left[\hat{Y}_{q+1}(q+1; \boldsymbol{\rho}) - \hat{Y}_{q+1}(k; \boldsymbol{\rho})\right] \qquad \text{(S10)}$$

where $\hat{Y}_{q+1}(x; \boldsymbol{\rho}) = \frac{\sum_{j=1}^{N} Y_{q+1,j}(\mathbf{X}) I[X_j = x]}{\sum_{j=1}^{N} I[X_j = x]}$ for $x \in \{0, \dots, q+1\}$. That is, $\hat{Y}_{q+1}(x; \boldsymbol{\rho})$ is the average of observed outcomes for individuals assigned with $x$ under strategy $\boldsymbol{\rho}$. Under Assumpstions 1 and 2,

$$E\left[\widehat{\delta_{q+1}}^{D}(\boldsymbol{\phi}, \boldsymbol{\rho})\right] = \delta_{q+1}^{D}(\boldsymbol{\phi}, \boldsymbol{\rho}).$$

*Proof:* First, we expand LHS:

$$E\left[\widehat{\delta_{q+1}}^{D}(\boldsymbol{\phi},\boldsymbol{\rho})\right] = E\left[N \cdot \sum_{k=0}^{q} \rho_k \cdot \left[\hat{Y}_{q+1}(q+1;\boldsymbol{\rho}) - \hat{Y}_{q+1}(k;\boldsymbol{\rho})\right]\right]$$

Linearity of expectation ⟶

$$= N \cdot \sum_{k=0}^{q}\left[\rho_k \cdot \left(E\left[\hat{Y}_{q+1}(q+1;\boldsymbol{\rho})\right] - E\left[\hat{Y}_{q+1}(k;\boldsymbol{\rho})\right]\right)\right] \quad \text{(S11).}$$

We largely repeat the proof in A.1 of Hudgens and Halloran[11] to evaluate $E\left[\hat{Y}_{q+1}(x;\boldsymbol{\rho})\right]$ for $x \in \{0, \dots, q+1\}$. Without loss of generality, let $x = 0$.

$$E\left[\hat{Y}_{q+1}(0;\boldsymbol{\rho})\right] = E\left[\frac{\sum_{j=1}^{N} Y_{q+1,j}(\mathbf{X}) I\left[X_j = 0\right]}{\sum_{j=1}^{N} I\left[X_j = 0\right]}\right]$$

Under Assumption 1, $K_0 \equiv N \cdot \rho_0$ is fixed by design, and under Assumption 2 (Consistency), we have:

$$E\left[\hat{Y}_{q+1}(0;\boldsymbol{\rho})\right] = \frac{1}{K_0}\sum_{j=1}^{N}\sum_{\mathbf{s}\in\mathcal{X}(N)} \Pr_{\boldsymbol{\rho}}(\mathbf{X} = \mathbf{s}) Y_{q+1,j}(\mathbf{s}) I\left[x_j = 0\right].$$

Any $\mathbf{s}$ such that $x_j \neq 0$ does not contribute to the summation, so that we can write:

$$E\left[\hat{Y}_{q+1}(0;\boldsymbol{\rho})\right]$$

$$= \tfrac{1}{K_0}\sum_{j=1}^{N}\sum_{\boldsymbol{\omega}\in\mathcal{X}(N-1)} \Pr_{\boldsymbol{\rho}}\left(\mathbf{X}_{-j} = \boldsymbol{\omega}, X_j = 0\right) Y_{q+1,j}\left(\mathbf{x}_{-j} = \boldsymbol{\omega}, x_j = 0\right)$$

$$= \tfrac{1}{K_0}\sum_{j=1}^{N}\sum_{\boldsymbol{\omega}\in\mathcal{X}(N-1)} \Pr_{\boldsymbol{\rho}}\left(\mathbf{X}_{-j} = \boldsymbol{\omega} | X_j = 0\right) \Pr_{\boldsymbol{\rho}}\left(X_j = 0\right) Y_{q+1,j}\left(\mathbf{x}_{-j} = \boldsymbol{\omega}, x_j = 0\right).$$

Under Assumption 1, $\Pr_{\boldsymbol{\rho}}\left(X_j = 0\right) = \rho_0 = K_0/N$, implying

$$E\left[\hat{Y}_{q+1}(0;\boldsymbol{\rho})\right] = \frac{1}{N}\sum_{j=1}^{N}\sum_{\boldsymbol{\omega}\in\mathcal{X}(N-1)} \Pr_{\boldsymbol{\rho}}\left(\mathbf{X}_{-j} = \boldsymbol{\omega} | X_j = 0\right) Y_{q+1,j}\left(\mathbf{x}_{-j} = \boldsymbol{\omega}, x_j = 0\right)$$

Recall $\bar{Y}_{q+1,j}(0;\boldsymbol{\rho}) \equiv \sum_{\boldsymbol{\omega}\in\mathcal{X}(N-1)} \Pr_{\boldsymbol{\rho}}\left(\mathbf{X}_{-j} = \boldsymbol{\omega} | X_j = x\right) Y_{q+1,j}\left(\mathbf{x}_{-j} = \boldsymbol{\omega}, x_j = 0\right) \longrightarrow$

$= \frac{1}{N} \sum_{j=1}^{N} \overline{Y}_{q+1,j}(0; \boldsymbol{\rho})$

Recall $\overline{Y}_{q+1}(0; \boldsymbol{\rho}) = \dfrac{\sum_{j=1}^{N} \overline{Y}_{q+1,j}(0; \boldsymbol{\rho})}{N} \longrightarrow$

$= \overline{Y}_{q+1}(0; \boldsymbol{\rho})$.

Recall equation (S11):

$$E\left[\widehat{\delta_{q+1}}^{D}(\boldsymbol{\phi}, \boldsymbol{\rho})\right] = N \cdot \sum_{k=0}^{q} \left[\rho_k \cdot \left(E\left[\hat{Y}_{q+1}(q+1; \boldsymbol{\rho})\right] - E\left[\hat{Y}_{q+1}(k; \boldsymbol{\rho})\right]\right)\right]$$

Substituting from above $\longrightarrow$

$$= N \cdot \sum_{k=0}^{q} \rho_k \left[\overline{Y}_{q+1}(q+1; \boldsymbol{\rho}) - \overline{Y}_{q+1}(k; \boldsymbol{\rho})\right]$$

Recall equation (S1) $\longrightarrow$

$= \delta_{q+1}^{D}(\boldsymbol{\phi}, \boldsymbol{\rho})$

**eAppendix 6. Extending the hazard difference estimator for vaccination at more than two time points**

**1 Hazard difference estimator for quantifying directly averted outcomes**

The hazard difference estimator for strategy $\boldsymbol{\rho} = \{q+1, d; 0, \dots, q\}$ for $q \in \mathbb{N}$ can be defined as follows:

$$\widetilde{\delta_{q+1}}^{D*}(\boldsymbol{\phi}, \boldsymbol{\rho}) = \sum_{l=1}^{q+1} \left[ \widehat{N}_{l-1}^{v}(\boldsymbol{\rho}) \cdot \left( \hat{h}_l^u(\boldsymbol{\rho}) - \hat{h}_l^v(\boldsymbol{\rho}) \right) \right]$$

where $\widehat{N}_{l-1}^{v}(\boldsymbol{\rho}) = N \sum_{k=0}^{l-1} \rho_k \left(1 - \hat{Y}_{l-1}(k; \boldsymbol{\rho})\right)$, $\hat{h}_l^u(\boldsymbol{\rho}) = \frac{\sum_{k=l}^{q+1} \rho_k \Delta \hat{Y}_l(k; \boldsymbol{\rho})}{\sum_{k=l}^{q+1} \rho_k (1 - \hat{Y}_{l-1}(k; \boldsymbol{\rho}))}$, and $\hat{h}_l^v(\boldsymbol{\rho}) = \frac{\sum_{k=0}^{l-1} \rho_k \Delta \hat{Y}_l(k; \boldsymbol{\rho})}{\sum_{k=0}^{l-1} \rho_k (1 - \hat{Y}_{l-1}(k; \boldsymbol{\rho}))}$.

**2 Bias of hazard difference estimator relative to the causal estimand**

**2.1 Analytical comparison**

**2.1.1 Bias of hazard difference estimator for an arbitrary number of vaccination times**

To examine the bias of $\widetilde{\delta_{q+1}}^{D*}(\boldsymbol{\phi}, \boldsymbol{\rho})$ relative to $\delta_{q+1}^{D}(\boldsymbol{\phi}, \boldsymbol{\rho})$, we first expand $E[\widetilde{\delta_{q+1}}^{D*}(\boldsymbol{\phi}, \boldsymbol{\rho})]$ as follows:

$$E\left[\widetilde{\delta_{q+1}}^{D*}(\boldsymbol{\phi}, \boldsymbol{\rho})\right] = E\left[\sum_{l=1}^{q+1} \left[ \widehat{N}_{l-1}^{v}(\boldsymbol{\rho}) \cdot \left( \hat{h}_l^u(\boldsymbol{\rho}) - \hat{h}_l^v(\boldsymbol{\rho}) \right) \right]\right]$$

$$= E\left[\sum_{l=1}^{q+1} \left[ N \cdot \sum_{k=0}^{l-1} \rho_k \left(1 - \hat{Y}_{l-1}(k; \boldsymbol{\rho})\right) \cdot \left( \frac{\sum_{k=l}^{q+1} \rho_k \Delta \hat{Y}_l(k; \boldsymbol{\rho})}{\sum_{k=l}^{q+1} \rho_k (1 - \hat{Y}_{l-1}(k; \boldsymbol{\rho}))} - \frac{\sum_{k=0}^{l-1} \rho_k \Delta \hat{Y}_l(k; \boldsymbol{\rho})}{\sum_{k=0}^{l-1} \rho_k (1 - \hat{Y}_{l-1}(k; \boldsymbol{\rho}))} \right) \right]\right]$$

$$= N \left\{ E\left[ \sum_{l=1}^{q+1} \left[ \left[ \sum_{k=0}^{l-1} \rho_k \left(1 - \hat{Y}_{l-1}(k; \boldsymbol{\rho})\right) \right] \cdot \frac{\sum_{k=l}^{q+1} \rho_k \Delta \hat{Y}_l(k; \boldsymbol{\rho})}{\sum_{k=l}^{q+1} \rho_k \left(1 - \hat{Y}_{l-1}(k; \boldsymbol{\rho})\right)} - \sum_{k=0}^{l-1} \rho_k \Delta \hat{Y}_l(k; \boldsymbol{\rho}) \right] \right] \right\}$$

Linearity of expectation ⟶

$$= N \left\{ \sum_{l=1}^{q+1} E\left[ \left[ \sum_{k=0}^{l-1} \rho_k \left(1 - \hat{Y}_{l-1}(k; \boldsymbol{\rho})\right) \right] \cdot \frac{\sum_{k=l}^{q+1} \rho_k \Delta \hat{Y}_l(k; \boldsymbol{\rho})}{\sum_{k=l}^{q+1} \rho_k \left(1 - \hat{Y}_{l-1}(k; \boldsymbol{\rho})\right)} \right] - \sum_{l=1}^{q+1} \sum_{k=0}^{l-1} E[\rho_k \Delta \hat{Y}_l(k; \boldsymbol{\rho})] \right\}$$

By a proof parallel with that of Theorem S2, we have:

$$E\left[\widehat{\delta_{q+1}}^{D*}(\boldsymbol{\phi},\boldsymbol{\rho})\right]=N\left\{\sum_{l=1}^{q+1}E\left[\left[\sum_{k=0}^{l-1}\rho_k\left(1-\hat{Y}_{l-1}(k;\boldsymbol{\rho})\right)\right]\cdot\frac{\sum_{k=l}^{q+1}\rho_k\Delta\hat{Y}_l(k;\boldsymbol{\rho})}{\sum_{k=l}^{q+1}\rho_k\left(1-\hat{Y}_{l-1}(k;\boldsymbol{\rho})\right)}\right]-\sum_{l=1}^{q+1}\sum_{k=0}^{l-1}\rho_k\Delta\bar{Y}_l(k;\boldsymbol{\rho})\right\}$$

Rearranging terms ⟶

$$=N\left\{\sum_{l=1}^{q+1}E\left[\left(\sum_{k=0}^{l-1}\rho_k\right)\cdot\frac{\frac{\sum_{k=0}^{l-1}\rho_k\left(1-\hat{Y}_{l-1}(k;\boldsymbol{\rho})\right)}{\sum_{k=0}^{l-1}\rho_k}}{\frac{\sum_{k=l}^{q+1}\rho_k\left(1-\hat{Y}_{l-1}(k;\boldsymbol{\rho})\right)}{\sum_{k=l}^{q+1}\rho_k}}\cdot\frac{\sum_{k=l}^{q+1}\rho_k\Delta\hat{Y}_l(k;\boldsymbol{\rho})}{\sum_{k=l}^{q+1}\rho_k}\right]-\sum_{l=1}^{q+1}\sum_{k=0}^{l-1}\rho_k\Delta\bar{Y}_l(k;\boldsymbol{\rho})\right\}\qquad\text{(S12)}$$

If $\frac{\frac{\sum_{k=0}^{l-1}\rho_k\left(1-\hat{Y}_{l-1}(k;\boldsymbol{\rho})\right)}{\sum_{k=0}^{l-1}\rho_k}}{\frac{\sum_{k=l}^{q+1}\rho_k\left(1-\hat{Y}_{l-1}(k;\boldsymbol{\rho})\right)}{\sum_{k=l}^{q+1}\rho_k}}=1$ (i.e., vaccination has no effect), then we can write equation (S12) as:

$$E\left[\widehat{\delta_{q+1}}^{D*}(\boldsymbol{\phi},\boldsymbol{\rho})\right]=N\left\{\sum_{l=1}^{q+1}E\left[\left(\sum_{k=0}^{l-1}\rho_k\right)\cdot\frac{\sum_{k=l}^{q+1}\rho_k\Delta\hat{Y}_l(k;\boldsymbol{\rho})}{\sum_{k=l}^{q+1}\rho_k}\right]-\sum_{l=1}^{q+1}\sum_{k=0}^{l-1}\rho_k\Delta\hat{Y}_l(k;\boldsymbol{\rho})\right\}$$

By a proof parallel with that of Theorem S2, we have:

$$E\left[\widehat{\delta_{q+1}}^{D*}(\boldsymbol{\phi},\boldsymbol{\rho})\right]=N\left\{\sum_{l=1}^{q+1}\left(\sum_{k=0}^{l-1}\rho_k\right)\cdot\frac{\sum_{k=l}^{q+1}\rho_k\Delta\bar{Y}_l(k;\boldsymbol{\rho})}{\sum_{k=l}^{q+1}\rho_k}-\sum_{l=1}^{q+1}\sum_{k=0}^{l-1}\rho_k\Delta\bar{Y}_l(k;\boldsymbol{\rho})\right\}$$

Recall equation[S4] ⟶

$$=\delta_{q+1}^{D}(\boldsymbol{\phi},\boldsymbol{\rho}).$$

However, if $\frac{\frac{\sum_{k=0}^{l-1}\rho_k\left(1-\hat{Y}_{l-1}(k;\boldsymbol{\rho})\right)}{\sum_{k=0}^{l-1}\rho_k}}{\frac{\sum_{k=l}^{q+1}\rho_k\left(1-\hat{Y}_{l-1}(k;\boldsymbol{\rho})\right)}{\sum_{k=l}^{q+1}\rho_k}}\neq 1$, then $E\left[\widehat{\delta_{q+1}}^{D*}(\boldsymbol{\phi},\boldsymbol{\rho})\right]\neq\delta_{q+1}^{D}(\boldsymbol{\phi},\boldsymbol{\rho})$. That is, $\widehat{\delta_{q+1}}^{D*}(\boldsymbol{\phi},\boldsymbol{\rho})$ is a biased estimator if the survival among vaccinated individuals is different from that among unvaccinated individuals. In an ideal RCT, the differential survival is due to a non-null effect of vaccination.

### 2.1.2 An example: vaccination at two time points

From equation (S12), we have:

$$E\left[\widehat{\delta_2}^{D*}(\boldsymbol{\phi},\boldsymbol{\rho})\right]=E\left[\hat{N}_0^v(\boldsymbol{\rho})\left(\hat{h}_1^u(\boldsymbol{\rho})-\hat{h}_1^v(\boldsymbol{\rho})\right)+\hat{N}_1^v(\boldsymbol{\rho})\left(\hat{h}_2^u(\boldsymbol{\rho})-\hat{h}_2^v(\boldsymbol{\rho})\right)\right]$$

$$= N\left\{\sum_{l=1}^{2} E\left[\left(\sum_{k=0}^{l-1}\rho_k\right)\cdot\frac{\frac{\sum_{k=0}^{l-1}\rho_k\left(1-\hat{Y}_{l-1}(k;\boldsymbol{\rho})\right)}{\sum_{k=0}^{l-1}\rho_k}}{\frac{\sum_{k=l}^{2}\rho_k\left(1-\hat{Y}_{l-1}(k;\boldsymbol{\rho})\right)}{\sum_{k=l}^{2}\rho_k}}\cdot\frac{\sum_{k=l}^{2}\rho_k\Delta\hat{Y}_l(k;\boldsymbol{\rho})}{\sum_{k=l}^{2}\rho_k}\right]-\sum_{l=1}^{2}\sum_{k=0}^{l-1}\rho_k\Delta\bar{Y}_l(k;\boldsymbol{\rho})\right\}$$

$$= N\left\{E\left[\rho_0\cdot\frac{\frac{\rho_0\left(1-\hat{Y}_0(0;\boldsymbol{\rho})\right)}{\rho_0}}{\frac{\rho_1\left(1-\hat{Y}_0(1;\boldsymbol{\rho})\right)+\rho_2\left(1-\hat{Y}_0(2;\boldsymbol{\rho})\right)}{\rho_1+\rho_2}}\cdot\frac{\rho_1\Delta\hat{Y}_1(1;\boldsymbol{\rho})+\rho_2\Delta\hat{Y}_1(2;\boldsymbol{\rho})}{\rho_1+\rho_2}\right]\right.$$

$$+E\left[(\rho_0+\rho_1)\cdot\frac{\frac{\rho_0\left(1-\hat{Y}_1(0;\boldsymbol{\rho})\right)+\rho_1\left(1-\hat{Y}_1(1;\boldsymbol{\rho})\right)}{\rho_0+\rho_1}}{\frac{\rho_2\left(1-\hat{Y}_0(2;\boldsymbol{\rho})\right)}{\rho_2}}\cdot\frac{\rho_2\Delta\hat{Y}_2(2;\boldsymbol{\rho})}{\rho_2}\right]$$

$$\left.-\left[\rho_0\Delta\bar{Y}_1(0;\boldsymbol{\rho})+\left(\rho_0\Delta\bar{Y}_2(0;\boldsymbol{\rho})+\rho_1\Delta\bar{Y}_2(1;\boldsymbol{\rho})\right)\right]\right\}$$

since $\hat{Y}_0 \equiv 0$ $\longrightarrow$

$$= N\left\{E\left[\rho_0\cdot\frac{\rho_1\Delta\hat{Y}_1(1;\boldsymbol{\rho})+\rho_2\Delta\hat{Y}_1(2;\boldsymbol{\rho})}{\rho_1+\rho_2}\right]+E\left[(\rho_0+\rho_1)\cdot\frac{\rho_0\left(1-\hat{Y}_1(0;\boldsymbol{\rho})\right)+\rho_1\left(1-\hat{Y}_1(1;\boldsymbol{\rho})\right)}{(\rho_0+\rho_1)\left(1-\hat{Y}_1(2;\boldsymbol{\rho})\right)}\cdot\Delta\hat{Y}_2(2;\boldsymbol{\rho})\right]\right.$$

$$\left.-\left[\rho_0\Delta\bar{Y}_1(0;\boldsymbol{\rho})+\left(\rho_0\Delta\bar{Y}_2(0;\boldsymbol{\rho})+\rho_1\Delta\bar{Y}_2(1;\boldsymbol{\rho})\right)\right]\right\}$$

By a proof parallel with that of Theorem S2, we have:

$$= N\left\{\rho_0\cdot\frac{\rho_1\Delta\bar{Y}_1(1;\boldsymbol{\rho})+\rho_2\Delta\bar{Y}_1(2;\boldsymbol{\rho})}{\rho_1+\rho_2}+E\left[(\rho_0+\rho_1)\cdot\frac{\rho_0\left(1-\hat{Y}_1(0;\boldsymbol{\rho})\right)+\rho_1\left(1-\hat{Y}_1(1;\boldsymbol{\rho})\right)}{(\rho_0+\rho_1)\left(1-\hat{Y}_1(2;\boldsymbol{\rho})\right)}\cdot\Delta\hat{Y}_2(2;\boldsymbol{\rho})\right]\right.$$

$$\left.-\left[\rho_0\Delta\bar{Y}_1(0;\boldsymbol{\rho})+\left(\rho_0\Delta\bar{Y}_2(0;\boldsymbol{\rho})+\rho_1\Delta\bar{Y}_2(1;\boldsymbol{\rho})\right)\right]\right\}$$

rearranging terms $\rightarrow$

$$= N\left\{\rho_0\cdot\left(\frac{\rho_1\Delta\bar{Y}_1(1;\boldsymbol{\rho})+\rho_2\Delta\bar{Y}_1(2;\boldsymbol{\rho})}{\rho_1+\rho_2}-\Delta\bar{Y}_1(0;\boldsymbol{\rho})\right)+E\left[(\rho_0+\rho_1)\cdot\frac{\rho_0\left(1-\hat{Y}_1(0;\boldsymbol{\rho})\right)+\rho_1\left(1-\hat{Y}_1(1;\boldsymbol{\rho})\right)}{(\rho_0+\rho_1)\left(1-\hat{Y}_1(2;\boldsymbol{\rho})\right)}\cdot\Delta\hat{Y}_2(2;\boldsymbol{\rho})\right]\right.$$

$$\left.-\left(\rho_0\Delta\bar{Y}_2(0;\boldsymbol{\rho})+\rho_1\Delta\bar{Y}_2(1;\boldsymbol{\rho})\right)\right\} \qquad \text{(S13)}$$

From the preceding section, we conclude that $E\left[\widehat{\delta_2}^{D*}(\boldsymbol{\phi}, \boldsymbol{\rho})\right] \neq \delta_2^D(\boldsymbol{\phi}, \boldsymbol{\rho})$ if $\frac{\rho_0(1-\hat{Y}_1(0;\boldsymbol{\rho}))+\rho_1(1-\hat{Y}_1(1;\boldsymbol{\rho}))}{(\rho_0+\rho_1)(1-\hat{Y}_1(2;\boldsymbol{\rho}))} \neq 1$ (i.e., vaccination has a non-null effect).

## 2.2 Simulations

Here we consider strategy $\boldsymbol{\rho}'' = \{99,7; \mathbf{0.01}\}$, where 1% of the individuals are randomized to vaccination at $x \in \{0,1, \dots, 98\}$, or to no vaccination throughout (i.e., $x = 99$). To illustrate the bias of hazard difference estimator relative to the causal estimand, we simulate the epidemic under the same nine scenarios as described in **Table 1**. Compartments are stratified by $x \in \{0,1, \dots, 99\}$, such that within each stratum $x$, we specify a continuous-time SIRD model in term of continuous time $t$ as follows:

$$
\left.
\begin{aligned}
\frac{dS_x(t)}{dt} &= -\theta_x(t) \cdot \lambda(t) \cdot S_x(t) \\
\frac{dI_x(t)}{dt} &= \theta_x(t) \cdot \lambda(t) \cdot S_x(t) - \gamma \cdot I_x(t) \\
\frac{dR_x(t)}{dt} &= (1 - \kappa_x(t) \cdot \mu) \cdot \gamma \cdot I_x(t) \\
\frac{dD_x(t)}{dt} &= \kappa_x(t) \cdot \mu \cdot \gamma \cdot I_x(t)
\end{aligned}
\right\} \text{(S14)}
$$

where $\lambda(t) = \beta \cdot \frac{\sum_{x=0}^{q+1} I_x(t)}{N(t)}$ and $N(t) =$ the sum of all individuals alive at $t$. For simulation, the model was discretized to day time-steps from Day 0 (i.e., the beginning of interval 0) to Day 693 (i.e., the beginning of interval 99). All model parameters are the same as **Table S4**, except that we use time-varying parameter $\theta_x(t)$ to parameterize 1-VE$_{\text{inf}}$ / 100%:

$$
\theta_x(t) = \begin{cases} 1 & \text{if } t < \text{beginning of interval } x \\ \theta & \text{if } t \geq \text{beginning of interval } x \end{cases}
$$

for vaccination interval $x$ and time $t$, where $\theta_x(\cdot)$ can be 10% or 0%, depending on the scenario as outlined in **Table 1**. Similarly, we use time-varying parameter $\kappa_x(t)$ to parameterize 1-VE$_{\text{death}}$ / 100%:

$$\kappa_x(t) = \begin{cases} 1 & \text{if } t < \text{beginning of interval } x \\ \kappa & \text{if } t \geq \text{beginning of interval } x \end{cases}$$

for vaccination interval $x$ and time $t$, where $\kappa_x(\cdot)$ can be 10% or 0%, depending on the scenario as outlined in **Table 1**.

Initial values for the simulation are described in **Table S2**.

**TABLE S2** | Initial values of the simulations under strategy $\boldsymbol{\rho}'' = \{99,7; \mathbf{0.01}\}$.

| Variable | Initial condition(s) | Definition |
|---|---|---|
| $\boldsymbol{N}(\mathbf{0})$ | 300,000 | Number alive at baseline |
| $\boldsymbol{S_x}(\mathbf{0})$ | $(300{,}000 - 300) * 0.01$ | Number of susceptible individuals assigned to receive vaccination at the beginning of interval $x$ |
| $\boldsymbol{I_x}(\mathbf{0})$ | $300 * 0.01$ | Number of infectious individuals assigned to receive vaccination at the beginning of interval $x$ |
| $\boldsymbol{R_x}(\mathbf{0})$ | 0 | Number of recovered individuals assigned to receive vaccination at the beginning of interval $x$ |
| $\boldsymbol{D_x}(\mathbf{0})$ | 0 | Number of individuals assigned to receive vaccination at the beginning of interval $x$ who died due to infection |

**Figure S2** compares the expected value of hazard difference estimator and the causal estimand for directly averted infections and deaths under strategy $\boldsymbol{\rho}'' = \{99,7; \mathbf{0.01}\}$ before the beginning of Day 693, while **Table S3** shows the absolute and percentage biases. Consistent with the derivation above, the hazard difference estimator overestimates averted outcomes when $VE_{inf} > 0$, due to the lower survival of not-yet-vaccinated individuals compared to ever-vaccinated individuals (i.e., $\frac{\frac{\sum_{k=0}^{l-1} \rho_k''\left(1-\hat{Y}_{l-1}(k;\boldsymbol{\rho}'')\right)}{\sum_{k=0}^{l-1} \rho_k''}}{\frac{\sum_{k=l}^{q+1} \rho_k''\left(1-\hat{Y}_{l-1}(k;\boldsymbol{\rho}'')\right)}{\sum_{k=l}^{q+1} \rho_k''}} > 1$ in equation [S12]).

**TABLE S3** | Absolute and percentage bias of the hazard difference estimator relative to the causal estimand (reference) for averted outcomes under strategy $\boldsymbol{\rho}''$ under scenarios varied by infection-fatality rate (IFR), vaccine efficacy against infection ($VE_{inf}$) and vaccine efficacy against death given infection ($VE_{death}$), as outlined in **Table 1**.

| | **Scenario 1** | **Scenario 2** | **Scenario 3** | **Scenario 4** | **Scenario 5** | **Scenario 6** | **Scenario 7** | **Scenario 8** | **Scenario 9** |
|---|---|---|---|---|---|---|---|---|---|
| | IFR: 1%, $VE_{inf}$ = 90%, $VE_{death}$ = 0% | IFR: 10%, $VE_{inf}$ = 90%, $VE_{death}$= 0% | IFR: 100%, $VE_{inf}$ = 90%, $VE_{death}$ = 0% | IFR: 1%, $VE_{inf}$ = 0%, $VE_{death}$ = 90% | IFR: 10%, $VE_{inf}$ = 0%, $VE_{death}$= 90% | IFR: 100%, $VE_{inf}$ = 0%, $VE_{death}$ = 90% | IFR: 1%, $VE_{inf}$ = 90%, $VE_{death}$ = 90% | IFR: 10%, $VE_{inf}$ = 90%, $VE_{death}$ = 90% | IFR: 100%, $VE_{inf}$ = 90%, $VE_{death}$ = 90% |
| **Averted infections** [a] | 25930.18 (191.06%) | 27362.86 (204.35%) | 77260.67 (1883.18%) | 0.00 | 0.00 | 0.00 | 25926.36 (191.03%) | 27319.99 (203.97%) | 62417.26 (931.8%) |
| **Averted deaths** [a] | 0.57 (0.42%) | 60.81 (4.54%) | 71795.51 (1749.97%) | 0.76 (0.35%) | 81.7 (3.74%) | 164255.82 (763.85%) | 0.83 (0.35%) | 89 (3.77%) | 173931.79 (769.03%) |

[a] Percentage differences were calculated using values rounded to nine decimal places, to avoid uninterpretable, extremely small non-zero values (absolute value $< 10^{-9}$). All values in the table are rounded to two decimal places.

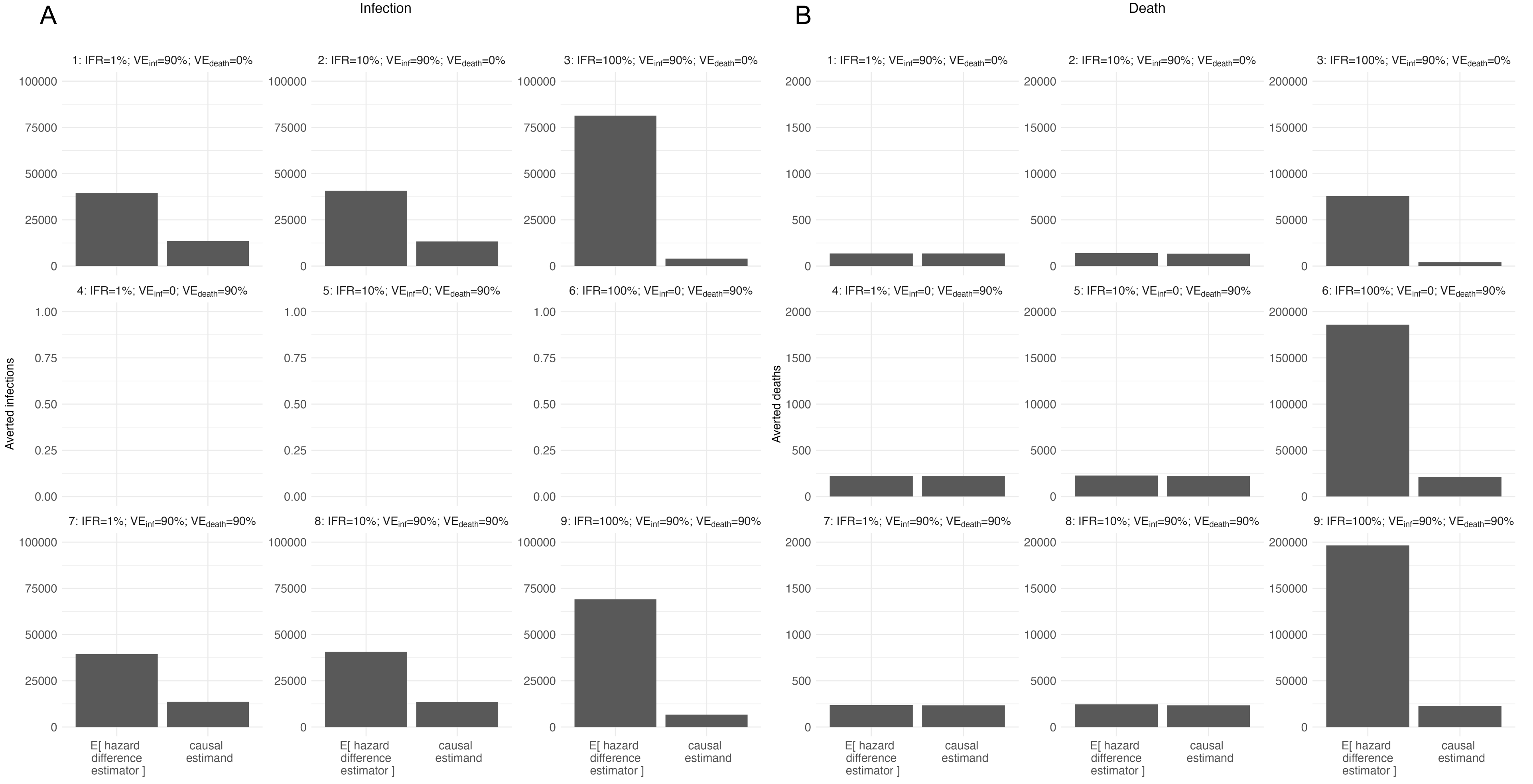

**FIGURE S2** | Infections (A) and deaths (B) directly averted by strategy $\boldsymbol{\rho''} = \{99{,}7; \mathbf{0.01}\}$ under scenarios varied by infection-fatality rate (IFR), vaccine efficacy against infection ($VE_{inf}$) and vaccine efficacy against death given infection ($VE_{death}$), as outlined in **Table 1**.

**eAppendix 7. An alternative estimator for quantifying directly averted outcomes using RCT data aggregated by vaccination status**

Here, we show that, for averted outcomes, the causal estimand can be identified with an alternative unbiased estimator using data aggregated by vaccination status through an expression similar (but not identical) to that of the hazard difference estimator. While this estimator may appear unnecessary for RCTs, where data are disaggregated by vaccination time and equation (S10) can be directly applied, it has important implications for empirical analyses that rely on aggregated data from national vaccine systems, as discussed further in **eAppendix 13**.

*Theorem S3 (Unbiasedness of an alternative estimator for averted outcomes).* Let $\boldsymbol{\rho} = \{q + 1, d; \rho_0, \dots, \rho_q\}$, where $\sum_{x=0}^{q+1} \rho_x = 1$, and $\boldsymbol{\phi} = \{q + 1, d; \mathbf{0}\}$ for $q \in \mathbb{N}$ and $d \in \mathbb{Z}^+$. Let

$$\widehat{\delta_{q+1}}^{D'}(\boldsymbol{\phi}, \boldsymbol{\rho}) = \sum_{l=1}^{q+1} \left( \frac{\sum_{k=0}^{l-1} \rho_k}{\sum_{k=l}^{q+1} \rho_k} \widehat{N}_{l-1}^{u}(\boldsymbol{\rho}) \hat{h}_l^{u}(\boldsymbol{\rho}) - \widehat{N}_{l-1}^{v}(\boldsymbol{\rho}) \hat{h}_l^{v}(\boldsymbol{\rho}) \right). \qquad \text{(S15)}$$

where $\widehat{N}_{l-1}^{u}(\boldsymbol{\rho}) = N \sum_{k=l}^{q+1} \rho_k \left(1 - \hat{Y}_{l-1}(k; \boldsymbol{\rho})\right)$, $\widehat{N}_{l-1}^{v}(\boldsymbol{\rho}) = N \sum_{k=0}^{l-1} \rho_k \left(1 - \hat{Y}_{l-1}(k; \boldsymbol{\rho})\right)$, $\hat{h}_l^{u}(\boldsymbol{\rho}) = \frac{\sum_{k=l}^{q+1} \rho_k \Delta \hat{Y}_l(k; \boldsymbol{\rho})}{\sum_{k=l}^{q+1} \rho_k \left(1 - \hat{Y}_{l-1}(k; \boldsymbol{\rho})\right)}$, and $\hat{h}_l^{v}(\boldsymbol{\rho}) = \frac{\sum_{k=0}^{l-1} \rho_k \Delta \hat{Y}_l(k; \boldsymbol{\rho})}{\sum_{k=0}^{l-1} \rho_k \left(1 - \hat{Y}_{l-1}(k; \boldsymbol{\rho})\right)}$.

Under Assumptions 1 and 2 from eAppendix 5, $E\left[\widehat{\delta_{q+1}}^{D'}(\boldsymbol{\phi}, \boldsymbol{\rho})\right] = \delta_{q+1}^{D}(\boldsymbol{\phi}, \boldsymbol{\rho})$.

*Proof*:

$$E[\widehat{\delta_{q+1}}^{D'}(\boldsymbol{\phi}, \boldsymbol{\rho})] = E\left[\sum_{l=1}^{q+1} \left( \frac{\sum_{k=0}^{l-1} \rho_k}{\sum_{k=l}^{q+1} \rho_k} \widehat{N}_{l-1}^{u}(\boldsymbol{\rho}) \hat{h}_l^{u}(\boldsymbol{\rho}) - \widehat{N}_{l-1}^{v}(\boldsymbol{\rho}) \hat{h}_l^{v}(\boldsymbol{\rho}) \right)\right]$$

Substituting $\widehat{N}_{l-1}^{u}(\boldsymbol{\rho}), \hat{h}_l^{u}(\boldsymbol{\rho}), \widehat{N}_{l-1}^{v}(\boldsymbol{\rho})$, and $\hat{h}_l^{v}(\boldsymbol{\rho})$ from above $\longrightarrow$

$$= \sum_{l=1}^{q+1} E\left[\left( N \cdot \frac{\sum_{k=0}^{l-1} \rho_k}{\sum_{k=l}^{q+1} \rho_k} \cdot \sum_{k=l}^{q+1} \rho_k \left(1 - \hat{Y}_{l-1}(k; \boldsymbol{\rho})\right) \cdot \frac{\sum_{k=l}^{q+1} \rho_k \Delta \hat{Y}_l(k; \boldsymbol{\rho})}{\sum_{k=l}^{q+1} \rho_k \left(1 - \hat{Y}_{l-1}(k; \boldsymbol{\rho})\right)} - N \cdot \sum_{k=0}^{l-1} \rho_k \left(1 - \hat{Y}_{l-1}(k; \boldsymbol{\rho})\right) \cdot \frac{\sum_{k=0}^{l-1} \rho_k \Delta \hat{Y}_l(k; \boldsymbol{\rho})}{\sum_{k=0}^{l-1} \rho_k \left(1 - \hat{Y}_{l-1}(k; \boldsymbol{\rho})\right)} \right)\right]$$

$$= N\sum_{l=1}^{q+1}\left(E\left[\frac{\sum_{k=0}^{l-1}\rho_k}{\sum_{k=l}^{q+1}\rho_k}\cdot\sum_{k=l}^{q+1}\rho_k\Delta\hat{Y}_l(k;\boldsymbol{\rho})\right]-E\left[\sum_{k=0}^{l-1}\rho_k\Delta\hat{Y}_l(k;\boldsymbol{\rho})\right]\right).$$

By a proof parallel with that of Theorem S2, we have:

$$E[\widehat{\delta_{q+1}}^{D'}(\boldsymbol{\phi},\boldsymbol{\rho})] = N\sum_{l=1}^{q+1}\left(\frac{\sum_{k=0}^{l-1}\rho_k}{\sum_{k=l}^{q+1}\rho_k}\cdot\sum_{k=l}^{q+1}\rho_k\Delta\bar{Y}_l(k;\boldsymbol{\rho})-\sum_{k=0}^{l-1}\rho_k\Delta\bar{Y}_l(k;\boldsymbol{\rho})\right)$$

Recall equation [S4] ⟶

$$= \delta_{q+1}^{D}(\boldsymbol{\phi},\boldsymbol{\rho})$$

Note that under the null, $\frac{\sum_{k=0}^{l-1}\rho_k}{\sum_{k=l}^{q+1}\rho_k}\widehat{N}_{l-1}^{u}(\boldsymbol{\rho}) = \widehat{N}_{l-1}^{v}(\boldsymbol{\rho})$ and equation (S15) can be written as $\sum_{l=1}^{q+1}\left[\widehat{N}_{l-1}^{v}(\boldsymbol{\rho})\left(\hat{h}_l^{u}(\boldsymbol{\rho})-\hat{h}_l^{v}(\boldsymbol{\rho})\right)\right]$, which reduces to the hazard difference estimator in **eAppendix 6**. Therefore, the same type of aggregated data required by the hazard difference estimator can be used by an unbiased estimator to quantify directly averted outcomes. However, when using aggregated data to identify the causal estimand in observational studies, we recognize that further procedures are needed to estimate $\rho_x$ for $x \in \{0, \ldots, q+1\}$ because there are no random assignments of **X**. We will further explore this in **eAppendix 13**.

## eAppendix 8. Parameters for simulations in the main text

**TABLE S4** | List of parameters for simulations

| Parameter | Value | Definition |
|---|---|---|
| $\boldsymbol{\beta}$ | 0.25 | The number of effective contacts made by a typical infectious individual per day |
| $\boldsymbol{\mu}$ | varied | Infection-fatality rate |
| $\boldsymbol{\theta}$ | varied | 1 – vaccine efficacy against infection (i.e., VE $_{\text{inf}}$ / 100%) |
| $\boldsymbol{\theta_1(t)}$ | 1 if $t$ < 60 days<br>$\theta$ if $t \geq$ 60 days | Same as above |
| $\boldsymbol{\kappa}$ | varied | 1 – vaccine efficacy against death given infection (i.e., VE $_{\text{death}}$ /100%) |
| $\boldsymbol{\kappa_1(t)}$ | 1 if $t$ < 60 days<br>$\kappa$ if $t \geq$ 60 days | Same as above |
| $\boldsymbol{\gamma}$ | 0.07 | Recovery rate per day |

**TABLE S5 |** List of initial conditions in the group under strategy $\boldsymbol{\rho}' = \{2,60; 0.2,0.3\}$

| Variable | Initial condition(s) | Definition |
|---|---|---|
| $\boldsymbol{N(0)}$ | 300,000 | Number alive at baseline |
| $\boldsymbol{S_2(0)}$ | $(300{,}000 - 300) * (1 - 0.2 - 0.3)$ | Number of susceptible individuals randomized to receive no vaccination |
| $\boldsymbol{S_1(0)}$ | $(300{,}000 - 300) * 0.3$ | Number of susceptible individuals randomized to receive on Day 60 |
| $\boldsymbol{S_0(0)}$ | $(300{,}000 - 300) * 0.2$ | Number of susceptible individuals randomized to receive vaccination at baseline |
| $\boldsymbol{I_2(0)}$ | $300 * (1 - 0.2 - 0.3)$ | Number of infectious individuals randomized to receive no vaccination |
| $\boldsymbol{I_1(0)}$ | $300 * 0.3$ | Number of infectious individuals randomized to receive vaccination on Day 60 |
| $\boldsymbol{I_0(0)}$ | $300 * 0.2$ | Number of infectious individuals randomized to receive vaccination at baseline |
| $\boldsymbol{R_2(0)}$ | 0 | Number of recovered individuals randomized to receive no vaccination |
| $\boldsymbol{R_1(0)}$ | 0 | Number of recovered individuals randomized to receive vaccination on Day 60 |
| $\boldsymbol{R_0(0)}$ | 0 | Number of recovered individuals randomized to receive vaccination at baseline |
| $\boldsymbol{D_2(0)}$ | 0 | Number of individuals randomized to receive no vaccination who died due to infection |
| $\boldsymbol{D_1(0)}$ | 0 | Number of individuals randomized to receive vaccination on Day 60 who died due to infection |
| $\boldsymbol{D_0(0)}$ | 0 | Number of individuals randomized to receive vaccination at baseline who died due to infection |

## eAppendix 9. Bias of the hazard difference estimator relative to the causal estimand on the absolute and relative scales

**TABLE S6** | Absolute and percentage bias of the hazard difference estimator relative to the causal estimand (reference) for averted outcomes under strategy $\boldsymbol{\rho}'$ under scenarios varied by infection-fatality rate (IFR), vaccine efficacy against infection ($VE_{inf}$) and vaccine efficacy against death given infection ($VE_{death}$), as outlined in **Table 1**.

| | **Scenario 1** | **Scenario 2** | **Scenario 3** | **Scenario 4** | **Scenario 5** | **Scenario 6** | **Scenario 7** | **Scenario 8** | **Scenario 9** |
|---|---|---|---|---|---|---|---|---|---|
| | IFR: 1%, $VE_{inf}$ = 90%, $VE_{death}$ = 0% | IFR: 10%, $VE_{inf}$ = 90%, $VE_{death}$= 0% | IFR: 100%, $VE_{inf}$ = 90%, $VE_{death}$ = 0% | IFR: 1%, $VE_{inf}$ = 0%, $VE_{death}$ = 90% | IFR: 10%, $VE_{inf}$ = 0%, $VE_{death}$= 90% | IFR: 100%, $VE_{inf}$ = 0%, $VE_{death}$ = 90% | IFR: 1%, $VE_{inf}$ = 90%, $VE_{death}$ = 90% | IFR: 10%, $VE_{inf}$ = 90%, $VE_{death}$ = 90% | IFR: 100%, $VE_{inf}$ = 90%, $VE_{death}$ = 90% |
| **Averted infections** [a] | 27139.84 (47.39%) | 28279.75 (49.63%) | 40903.49 (108.29%) | 0.00 | 0.00 | 0.00 | 27127.71 (47.37%) | 28151.4 (49.43%) | 39707.74 (89.09%) |
| **Averted deaths** [a] | 1.08 (0.19%) | 114.18 (2.04%) | 24399.47 (61.16%) | 1.09 (0.14%) | 116.4 (1.52%) | 38964.52 (51.95%) | 1.24 (0.13%) | 131.14 (1.38%) | 28154.09 (28.75%) |

[a] Percentage differences were calculated using values rounded to nine decimal places, to avoid uninterpretable, extremely small non-zero values (absolute value $< 10^{-9}$). All values in the table are rounded to two decimal places.

## eAppendix 10. Disease dynamics under each scenario

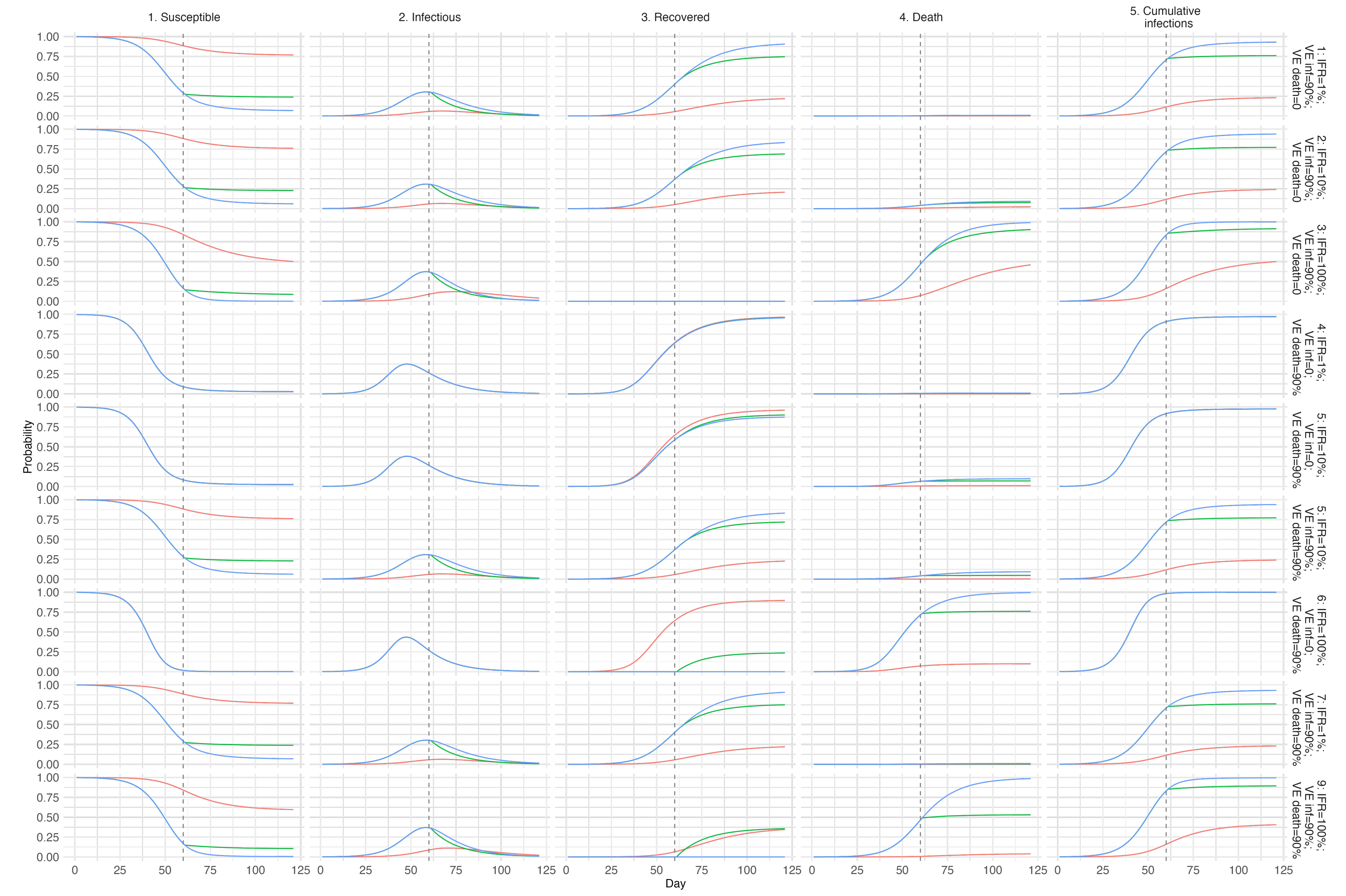

**FIGURE S3** | Disease dynamics under strategy $\boldsymbol{\rho}' = \{2,60; 0.2,0.3\}$ in scenarios varied by infection-fatality rate (IFR), vaccine efficacy against infection ($VE_{inf}$) and vaccine efficacy against death given infection ($VE_{death}$). Red line represents individuals assigned to vaccination at baseline, green for those assigned to vaccination at Day 60, and blue for those remaining unvaccinated throughout. Grey dashed line indicates Day 60.

**eAppendix 11. Scenarios for simulations by varying the number of effective contacts**

In **Figure 2B**, we varied IFR and showed that $\widehat{\delta_2}^{D*}(\boldsymbol{\phi}, \boldsymbol{\rho}')$ substantially overestimates averted deaths when IFR = 100%, due to the stark difference in relative survival between vaccinated and not-yet-vaccinated individuals. However, when IFR ≤ 10%, the overestimation is trivial due to the similar survival across groups. Here, instead of varying IFR, we vary the number of effective contacts ($\beta$) to explore the bias. Scenarios are specified in **Table S7**. All other parameters match the main analysis (**Table S4**), except for infection-fatality rate $\mu$, fixed at 0.1. Initial values also match the main analysis (**Table S5**).

**TABLE S7** | Scenarios varied by number of effective contacts ($\beta$), vaccine efficacy against infection ($VE_{inf}$) and vaccine efficacy against death given infection ($VE_{death}$)

| **Scenario** | $\beta$ | **Vaccine efficacy** |
|---|---|---|
| Scenario I | 0.15 | $VE_{inf}$ = 90%; $VE_{death}$ = 0% |
| Scenario II | 0.2 | $VE_{inf}$ = 90%; $VE_{death}$ = 0% |
| Scenario III | 0.25 | $VE_{inf}$ = 90%; $VE_{death}$ = 0% |
| Scenario IV | 0.15 | $VE_{inf}$ = 0%; $VE_{death}$ = 90% |
| Scenario V | 0.2 | $VE_{inf}$ = 0%; $VE_{death}$ = 90% |
| Scenario VI | 0.25 | $VE_{inf}$ = 0%; $VE_{death}$ = 90% |
| Scenario VII | 0.15 | $VE_{inf}$ = 90%; $VE_{death}$ = 90% |
| Scenario VIII | 0.2 | $VE_{inf}$ = 90%; $VE_{death}$ = 90% |
| Scenario IX | 0.25 | $VE_{inf}$ = 90%; $VE_{death}$ = 90% |

**Figure S4** compares the expected value of hazard difference estimator with the causal estimand for averted infections and deaths in Scenarios as specified in **Table S7**, while **Table S8** shows the absolute and percentage bias. As shown in **Figure S4** and **Table S8**, the hazard difference estimator overestimates averted infections in all Scenarios (except for averted infections when $VE_{inf}$ = 0%; Scenarios IV to VI), and the bias increases with $\beta$. This is because susceptibles (from infections) are preferentially depleted among unvaccinated individuals at a faster rate when $\beta$ is higher and $VE_{inf}$ = 90%.

**FIGURE S4** | Infections (A) and deaths (B) directly averted by strategy $\boldsymbol{\rho}' = \{2{,}60; 0.2{,}0.3\}$ among vaccinated individuals under different scenarios varied by number of effective contacts ($\beta$), vaccine efficacy against infection ($VE_{inf}$), and vaccine efficacy against death given infection ($VE_{death}$), as specified in **Table S7**

**TABLE S8** | Absolute and percentage bias of the hazard difference estimator relative to the causal estimand (reference) for averted outcomes under strategy $\boldsymbol{\rho}' = \{2{,}60; 0.2{,}0.3\}$ under scenarios varied by the number of effective contacts ($\beta$), vaccine efficacy against infection ($VE_{inf}$) and vaccine efficacy against death given infection ($VE_{death}$) as specified in **Table S7**

| | **Scenario I** | **Scenario II** | **Scenario III** | **Scenario IV** | **Scenario V** | **Scenario VI** | **Scenario VII** | **Scenario VIII** | **Scenario IX** |
|---|---|---|---|---|---|---|---|---|---|
| | $\beta = 0.15$, $VE_{inf} = 90\%$, $VE_{death} = 0\%$ | $\beta = 0.2$, $VE_{inf} = 90\%$, $VE_{death} = 0\%$ | $\beta = 0.25$, $VE_{inf} = 90\%$, $VE_{death} = 0\%$ | $\beta = 0.15$, $VE_{inf} = 0\%$, $VE_{death} = 90\%$ | $\beta = 0.2$, $VE_{inf} = 0\%$, $VE_{death} = 90\%$ | $\beta = 0.25$, $VE_{inf} = 0\%$, $VE_{death} = 90\%$ | $\beta = 0.15$, $VE_{inf} = 90\%$, $VE_{death} = 90\%$ | $\beta = 0.2$, $VE_{inf} = 90\%$, $VE_{death} = 90\%$ | $\beta = 0.25$, $VE_{inf} = 90\%$, $VE_{death} = 90\%$ |
| **Averted infections** [a] | 551.26 (2.1%) | 10575.04 (14.47%) | 28279.75 (49.63%) | 0.00 | 0.00 | 0.00 | 550.01 (2.1%) | 10526.5 (14.44%) | 28151.4 (49.43%) |
| **Averted deaths** [a] | 2.68 (0.13%) | 45.64 (0.68%) | 114.18 (2.04%) | 25.47 (0.29%) | 112.22 (1.18%) | 116.4 (1.52%) | 3.03 (0.12%) | 50.86 (0.56%) | 131.14 (1.38%) |

[a] Percentage differences were calculated using values rounded to nine decimal places, to avoid uninterpretable, extremely small non-zero values (absolute value $< 10^{-9}$). All values in the table are rounded to two decimal places.

**eAppendix 12. Simulations with more realistic parameter values**

In the main text, we considered extreme combinations of IFR, $VE_{inf}$, and $VE_{death}$ for illustrative purposes. Here, we consider more realistic parameter values estimated by previous studies on seasonal flu, measles, and COVID-19 (wild-type strain), as shown in **Table S9**. As in the main text, we simulate strategy $\boldsymbol{\rho}' = \{2,60; 0.2,0.3\}$, using the same model structure as in equation (8). Our objective is not to realistically model the dynamics of these epidemics—which would involve population heterogeneity, more complex infection characteristics (i.e., breakthrough infections, re-infections), waning immunity, vaccination at more time steps, and parameters that vary by time and context. Instead, we simply aim to test the robustness of our results across a broader and more realistic range of parameter values compared to those parameters used in the main text (as presented in **Table S4**).

**TABLE S9** | More realistic parameter values for seasonal flu, measles, and COVID-19 (wildtype). Definitions for the parameters are provided in **Table S4**. Model equation is shown in equation (8).

| Parameter | Seasonal flu | Measles | COVID-19 |
|---|---|---|---|
| $\boldsymbol{\beta}$[a] | 0.41 [15] | 1.26 [16] | 0.15 [17] |
| $\boldsymbol{\mu}$ | 3% [18] | 1.3% [19] | 0.11% [20] |
| $\boldsymbol{\theta}$ | 0.66 [21] | 0.05 [22] | 0.05 [23] |
| $\boldsymbol{\kappa}$ | 0.69 [24] | 0 [25] | 0.04 [26] |
| $\boldsymbol{\gamma}$ | 0.21 [15] | 0.07 [27] | 0.07 [28] |

[a] $\beta$ is obtained from applying the formula $\beta = R_0 \cdot \gamma$, where $R_0$ =basic reproduction number.

As **Figure S5** and **Table S10** shows, the hazard difference estimator substantially overestimates averted infections for measles due to high basic reproduction number ($R_0$=18) and high $VE_{inf}$. It also slightly overestimates averted infections for seasonal flu and COVID-19 (wild-type), given a low $R_0$ (i.e., $R_0 \leq 2.2$) and $VE_{inf} > 0$. The overestimation of averted deaths is trivial due to the low IFR ($\leq$ 3% for all pathogens).

For avertible outcomes (defined in **eAppendix 14**), the hazard difference estimator slightly underestimates avertible deaths for COVID-19. As shown later, this occurs because the hazard difference estimator neglects the term $\rho_1 \cdot [\Delta\bar{Y}_2(1;\boldsymbol{\rho}) - \Delta\bar{Y}_2(0;\boldsymbol{\rho})]$ in the causal estimand (equation [S20]). This term is positive for COVID-19 (i.e., the period incidence of death for individuals with $x = 1$ is higher than that for individuals with $x = 0$), such that hazard difference estimator underestimates the causal estimand. For measles, there is no bias for avertible deaths under the hazard difference estimator because $\Delta\bar{Y}_2(0;\boldsymbol{\rho}) = \Delta\bar{Y}_2(1;\boldsymbol{\rho}) = 0$ (See equation [S20] and compare it with equation [S24]).

**TABLE S10** | Absolute and percentage bias of the hazard difference estimator relative to the causal estimand (reference) for averted and avertible outcomes under strategy $\boldsymbol{\rho}' = (2{,}60; 0.2, 0.3)$ under more realistic parameter values for seasonal flu, measles, and COVID-19 (wild type strain)

| | **Seasonal flu** | **Measles** | **COVID-19 (wild type strain)** |
|---|---|---|---|
| **Averted infections** [a] | 939.98<br>(9.59%) | 12133.48<br>(46.69%) | 462.12<br>(1.94%) |
| **Averted deaths** [a] | 1.02<br>(0.15%) | 0.26<br>(0.03%) | 0.00<br>(0.00%) |
| **Avertible infections** [a] | 2524.04<br>(7.13%) | 3153.28<br>(3.03%) | 99.9<br>(0.31%) |
| **Avertible deaths** [a] | 39.49<br>(1.63%) | 0<br>(0%) | -0.16<br>(-0.53%) |

[a] Percentage differences were calculated using values rounded to nine decimal places, to avoid uninterpretable, extremely small non-zero values (absolute value $< 10^{-9}$). All values in the table are rounded to two decimal places.

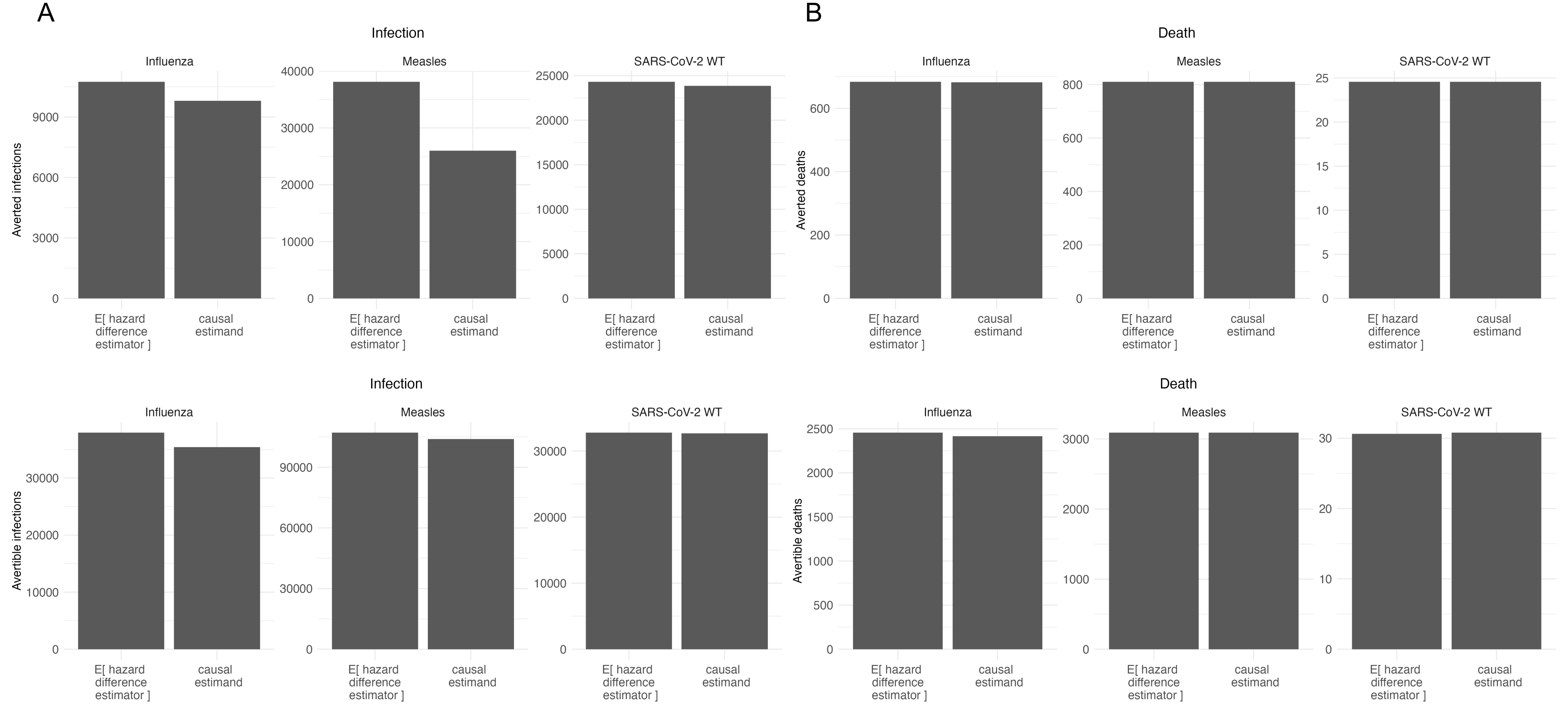

**FIGURE S5** | Infections (A) and deaths (B) directly averted by strategy $\boldsymbol{\rho}' = \{2,60; 0.2,0.3\}$ among vaccinated individuals under more realistic parameter values for seasonal flu, measles, and COVID-19 (wild type strain)

**eAppendix 13. Using observational data aggregated by vaccination status for estimation**

**eAppendix 7** defines an alternative unbiased estimator (equation [S15]) using RCT data aggregated by vaccination status. However, such an RCT is often not feasible in reality, and here we show how to identify the causal estimand using observational data aggregated by vaccination status (i.e., similar data structure as the national vaccine system[29]).

**1 Obtaining $\hat{\boldsymbol{\rho}}_x$ using observational data**

In the proposed RCT, we specify the strategy $\boldsymbol{\rho}$ and the proportions $\{\rho_0, \ldots, \rho_{q+1}\}$ by design. In an observational study, the investigators make no assignments, rather we only observe the time at which individuals are vaccinated (denoted as $\mathbf{X}^*$). In this case, the proportions $\rho_x$ for $x \in \{0, \ldots, q+1\}$, must be estimated from the data. However, we cannot assume $\rho_x = E\left[\frac{\sum_{j=1}^{N} I\left[X_j^* = x^*\right]}{N}\right]$ for $x^* \in \{0, \ldots, q+1\}$ due to immortal time bias (i.e., since individuals who die before interval 1 cannot receive vaccination, individuals with $x^* = 1$ are by definition "immortal" before interval 1).[30] Here, we discuss the procedures to estimate $\rho_x$ for $x \in \{0, \ldots, q+1\}$, addressing the immortal time bias. Note we assume the absence of confounding and additional selection bias at baseline and throughout follow-up. These assumptions are implausible in observational studies, but addressing all their violations is beyond the scope of this paper. Instead, this section focuses on identifying the causal estimand using observational data aggregated by vaccination status, addressing the immortal time bias that would otherwise exist even in the absence of confounding or additional selection bias.

Without loss of generality, consider discrete-time outcome Y for *death*. The SIRD-model in equation (S14) can be simplified into the following equation for discrete-time outcome Y:

$$N_{l+1}^{v}(\boldsymbol{\rho}) = N_{l}^{v}(\boldsymbol{\rho}) \cdot \left(1 - h_{l+1}^{v}(\boldsymbol{\rho})\right) + N \cdot \rho_{l+1} \cdot \prod_{k=0}^{l+1}\left(1 - h_{k}^{u}(\boldsymbol{\rho})\right) \qquad \text{(S16)}$$

where $N_l^v(\boldsymbol{\rho}) = N \sum_{k=0}^{l} \rho_k \left(1 - \bar{Y}_l(k; \boldsymbol{\rho})\right)$ is the number survived among vaccinated individuals by interval $l$, and $h_{l+1}^v(\boldsymbol{\rho}) = \frac{\sum_{k=0}^{l} \rho_k \Delta \bar{Y}_{l+1}(k;\boldsymbol{\rho})}{\sum_{k=0}^{l} \rho_k \left(1-\bar{Y}_l(k;\boldsymbol{\rho})\right)}$ (or $h_{l+1}^u(\boldsymbol{\rho}) = \frac{\sum_{k=l+1}^{q+1} \rho_k \Delta \bar{Y}_{l+1}(k;\boldsymbol{\rho})}{\sum_{k=l+1}^{q+1} \rho_k \left(1-\bar{Y}_l(k;\boldsymbol{\rho})\right)}$) is the hazard from interval $l$ to $l+1$ among vaccinated (or unvaccinated) individuals.

Therefore, $\rho_{l+1}$ can be estimated by:

$$\hat{\rho}_{l+1} = \frac{\widehat{N}_{l+1}^v(\boldsymbol{\rho}) - \widehat{N}_l^v(\boldsymbol{\rho}) \cdot \left(1 - \hat{h}_{l+1}^v(\boldsymbol{\rho})\right)}{N \cdot \prod_{k=0}^{l+1} \left(1 - \hat{h}_k^u(\boldsymbol{\rho})\right)}. \qquad \text{(S17)}$$

To illustrate this, **Table S11** shows a simulated dataset that is aggregated by vaccination status in a hypothetical observational study in the absence of confounding or additional selection bias. The epidemic is simulated under the (unobserved) strategy $\boldsymbol{\rho}' = \{2, d; 0.2, 0.3\}$ in Scenario 1 as described in **Table 1**.

**TABLE S11** | Dataset aggregated by vaccination status in a hypothetical observational study without confounding under unobserved strategy $\boldsymbol{\rho}' = \{2,60; 0.2,0.3\}$

| Interval ($\boldsymbol{k}$) | Observed number survived among vaccinated individuals ($\widehat{\boldsymbol{N}}_{\boldsymbol{k}}^{\boldsymbol{v}}(\boldsymbol{\rho}')$) [a] | Hazards of death among vaccinated individuals ($\widehat{\boldsymbol{h}}_{\boldsymbol{k+1}}^{\boldsymbol{v}}(\boldsymbol{\rho}')$) [b] | Observed number survived among not-yet-vaccinated individuals ($\widehat{\boldsymbol{N}}_{\boldsymbol{k}}^{\boldsymbol{u}}(\boldsymbol{\rho}')$) [a] | Hazards of death among not-yet-vaccinated individuals ($\widehat{\boldsymbol{h}}_{\boldsymbol{k+1}}^{\boldsymbol{u}}(\boldsymbol{\rho}')$) [b] |
|---|---|---|---|---|
| 0 | 60000 | 0.00060716 | 240000 | 0.00426391 |
| 1 | 149579.818 | 0.0026133 | 149360.413 | 0.00491036 |

[a] Number of individuals survived before the start of interval $k$ among those who are (un)vaccinated at the start of the that interval.
[b] Hazard for events occurring after the start of interval $k$ and before the start of interval $k+1$ (i.e., hazard leads survival by one interval).

Assuming that we know the baseline vaccination proportion $\rho_0' = 0.2$, we are interested in recovering $\rho_1'$ and $\rho_2'$. By substituting the data from **Table S11** into equation (S17), we have:

$$\hat{\rho}_1' = \frac{\widehat{N}_1^v(\boldsymbol{\rho}') - \widehat{N}_0^v(\boldsymbol{\rho}') \cdot \left(1 - \hat{h}_1^v(\boldsymbol{\rho}')\right)}{\widehat{N} \cdot \left(1 - \hat{h}_0^u(\boldsymbol{\rho}')\right) \cdot \left(1 - \hat{h}_1^u(\boldsymbol{\rho}')\right)} = \frac{149579.818 - 60000 \cdot (1 - 0.00060716)}{300000 \cdot 1 \cdot (1 - 0.00426391)} = 0.3 = {\rho'}_1$$

Then, we can easily obtain $\hat{\rho}_2' = 1 - 0.2 - 0.3 = 0.5$.

## 3 Identifying causal estimand using observational data

Lastly, we verify that, the causal estimand can be identified using observational data aggregated by vaccination status (subject to immortal time bias until corrected by equation [S17]) in the absence of confounding or additional selection bias. By having $\boldsymbol{\rho}' = \widehat{\boldsymbol{\rho}}'$ and substituting the data from **Table S11** into equation (S15), we have:

$$\widehat{\delta_{q+1}}^{D'}(\boldsymbol{\phi}, \boldsymbol{\rho}') = \sum_{l=1}^{2} \left( \frac{\sum_{k=0}^{l-1} \hat{\rho}'_k}{\sum_{k=l}^{q+1} \hat{\rho}'_k} \widehat{N}_{l-1}^{u}(\boldsymbol{\rho}')\hat{h}_l^{u}(\boldsymbol{\rho}') - \widehat{N}_{l-1}^{v}(\boldsymbol{\rho}')\hat{h}_l^{v}(\boldsymbol{\rho}') \right)$$

$$= \left[ \frac{\hat{\rho}'_0}{\hat{\rho}'_1 + \hat{\rho}'_2} \widehat{N}_0^{u}(\boldsymbol{\rho}')\hat{h}_1^{u}(\boldsymbol{\rho}') - \widehat{N}_0^{v}(\boldsymbol{\rho}')\hat{h}_1^{v}(\boldsymbol{\rho}') \right] + \left[ \frac{\hat{\rho}'_0 + \hat{\rho}'_1}{\hat{\rho}'_2} \widehat{N}_1^{u}(\boldsymbol{\rho}')\hat{h}_2^{u}(\boldsymbol{\rho}') - \widehat{N}_1^{v}(\boldsymbol{\rho}')\hat{h}_2^{v}(\boldsymbol{\rho}') \right]$$

$$= (\frac{0.2}{0.8} * 240000 * 0.00426391 - 60000.0 * 0.00060716) + (\frac{0.5}{0.5} * 149360.413 * 0.00491036 - 149579.818 * 0.0026133)$$

$$= 561.92146$$

Consider the simulated data from a hypothetical RCT where we have cumulative incidence data disaggregated by vaccination time (**Table S12**). By substituting the RCT data into equation (S10) (i.e., the unbiased estimator), we estimate the averted deaths to be $300000 * [0.2 * (0.00915333 - 0.00220958) + 0.3 * (0.00915333 - 0.00753893)] = 561.921$, which is the same as the result obtained from **Table S11**. Therefore, in the absence of confounding or additional selection bias, we can use observational data aggregated by vaccination status to estimate cumulative incidence difference estimand for averted outcomes.

**TABLE S12** | Dataset disaggregated by vaccination time in a hypothetical one-stage RCT under strategy $\boldsymbol{\rho}' = \{2,60; 0.2,0.3\}$

| Vaccination time ($x$) | $\boldsymbol{\rho}'_{\boldsymbol{x}}$ | Cumulative incidence for death $\hat{Y}_2(x; \boldsymbol{\rho}')$ |
|---|---|---|
| 0 | 0.2 | 0.00220958 |
| 1 | 0.3 | 0.00753893 |
| 2 | 0.5 | 0.00915333 |

**eAppendix 14. Outcomes directly avertible by vaccination**

In the main text, we define the causal estimand and estimators for quantifying outcomes directly averted by vaccination. Here, we propose the causal estimand and estimators for outcomes directly *avertible* by vaccination. The proposed causal estimand considers outcomes that could have been averted under full vaccination at baseline (denoted as $\boldsymbol{\psi} = (q+1, d; 1, \mathbf{0})$ for $q \in \mathbb{N}$), but were not averted given the particular vaccination strategy $\boldsymbol{\rho}$.

**1.1 Causal estimand for outcomes directly avertible by vaccination at a single time point**

Our previous framework[14] defined the causal estimand for quantifying avertible outcomes for vaccination at a single time point—that is, outcomes that could have been directly averted under full vaccination at baseline $\boldsymbol{\psi} = \{1, d; 1\}$ but were not averted given the particular strategy $\boldsymbol{\rho} = \{1, d; \rho_0\}$, where $\rho_1 = 1 - \rho_0$. In notation, the estimand is:

$$\delta_1^D(\boldsymbol{\psi}, \boldsymbol{\rho}) = N \cdot \rho_1 \cdot \left(\bar{Y}_1(1; \boldsymbol{\rho}) - \bar{Y}_1(0; \boldsymbol{\rho})\right) \qquad \text{(S18)}$$

**1.2 Causal estimand for outcomes directly avertible by vaccination at two time points**

Now, extending the estimand in equation (S18) for vaccination at two time points, we define the estimand to quantify outcomes that could have been directly averted under full vaccination at baseline $\boldsymbol{\psi} = \{2, d; 1{,}0\}$ but were not averted given the particular strategy $\boldsymbol{\rho} = \{2, d; \rho_0, \rho_1\}$. In notation, the estimand is:

$$\delta_2^D(\boldsymbol{\psi}, \boldsymbol{\rho}) = N \cdot \left[\rho_1\left(\bar{Y}_2(1; \boldsymbol{\rho}) - \bar{Y}_2(0; \boldsymbol{\rho})\right) + \rho_2\left(\bar{Y}_2(2; \boldsymbol{\rho}) - \bar{Y}_2(0; \boldsymbol{\rho})\right)\right] \qquad \text{(S19)}$$

$$= N \cdot \left[\left(\rho_1 \Delta\bar{Y}_1(1; \boldsymbol{\rho}) + \rho_2 \Delta\bar{Y}_1(2; \boldsymbol{\rho})\right) - (\rho_1 + \rho_2) \cdot \Delta\bar{Y}_1(0; \boldsymbol{\rho}) + \rho_1 \cdot \left(\Delta\bar{Y}_2(1; \boldsymbol{\rho}) - \Delta\bar{Y}_2(0; \boldsymbol{\rho})\right)\right.$$

$$\left. + \rho_2 \cdot \left(\Delta\bar{Y}_2(2; \boldsymbol{\rho}) - \Delta\bar{Y}_2(0; \boldsymbol{\rho})\right)\right]. \qquad \text{(S20)}$$

In words, equation (S19) takes the difference in cumulative incidence between vaccination at baseline (i.e., $\bar{Y}_2(0; \boldsymbol{\rho})$) and vaccination at later time $x \in \{1,2\}$ (i.e., $\bar{Y}_2(1; \boldsymbol{\rho})$ or $\bar{Y}_2(2; \boldsymbol{\rho})$), multiplied with number of individuals assigned to $x$=1 or 2 (i.e., $N \cdot \rho_1$ or $N \cdot \rho_2$), respectively, and sum over $x \in \{1,2\}$.

**1.3 Outcomes directly avertible by vaccination at an arbitrary number of vaccination times**

Extending the estimand in equation (S19) for vaccination at an arbitrary number of vaccination times, we define the estimand to quantify outcomes that could have been directly averted under full vaccination at baseline $\boldsymbol{\psi} = \{q+1, d; 1, \mathbf{0}\}$ but were not averted given the particular strategy $\boldsymbol{\rho} = \{q+1, d; \rho_0, \dots, \rho_q\}$ for $q \in \mathbb{N}$. In notation, the estimand is:

$$\delta_{q+1}^{D}(\boldsymbol{\psi}, \boldsymbol{\rho}) = N \cdot \sum_{k=1}^{q+1} \rho_k \cdot \left(\bar{Y}_{q+1}(k; \boldsymbol{\rho}) - \bar{Y}_{q+1}(0; \boldsymbol{\rho})\right). \quad \text{(S21)}$$

**2 Unbiased estimator**

Similar to the unbiased estimator for directly averted outcomes in equation (S10), directly avertible outcomes (i.e., $\delta_{q+1}^{D}(\boldsymbol{\psi}, \boldsymbol{\rho})$) can be identified by:

$$\widehat{\delta_{q+1}}^{D}(\boldsymbol{\psi}, \boldsymbol{\rho}) = N \cdot \sum_{k=1}^{q+1} \rho_k \cdot \left[\hat{Y}_{q+1}(k; \boldsymbol{\rho}) - \hat{Y}_{q+1}(0; \boldsymbol{\rho})\right] \quad \text{(S22)}$$

where $\hat{Y}_{q+1}(x; \boldsymbol{\rho}) = \frac{\sum_{j=1}^{N} Y_{q+1,j}(\mathbf{X}) I[X_j = x]}{\sum_{j=1}^{N} I[X_j = x]}$ for $x \in \{0, \dots, q+1\}$. We know that $\widehat{\delta_{q+1}}^{D}(\boldsymbol{\psi}, \boldsymbol{\rho})$ is unbiased based on a proof parallel with the proof of Theorem S2.

**3 Hazard difference estimator for outcomes directly avertible by vaccination**

Some empirical studies[31,32] have used the hazard difference estimator to estimate vaccine-avertible outcomes, which takes the form:

$$\widehat{\delta_{q+1}}^{D*}(\boldsymbol{\psi}, \boldsymbol{\rho}) = \sum_{l=1}^{q+1} \left[\hat{N}_{l-1}^{u}(\boldsymbol{\rho}) \cdot \left(\hat{h}_l^{u}(\boldsymbol{\rho}) - \hat{h}_l^{v}(\boldsymbol{\rho})\right)\right]$$

$$= \hat{N}_0^{u}(\boldsymbol{\rho}) \cdot \left(\hat{h}_1^{u}(\boldsymbol{\rho}) - \hat{h}_1^{v}(\boldsymbol{\rho})\right) + \sum_{l=2}^{q+1} \left[\hat{N}_{l-1}^{u}(\boldsymbol{\rho}) \cdot \left(\hat{h}_l^{u}(\boldsymbol{\rho}) - \hat{h}_l^{v}(\boldsymbol{\rho})\right)\right] \quad \text{(S23)}$$

where $\widehat{N}_{l-1}^{u}(\boldsymbol{\rho}) = N\sum_{k=l}^{q+1}\rho_k\left(1-\widehat{Y}_{l-1}(k;\boldsymbol{\rho})\right)$, $\widehat{h}_l^{u}(\boldsymbol{\rho}) = \frac{\sum_{k=l}^{q+1}\rho_k\Delta\widehat{Y}_l(k;\boldsymbol{\rho})}{\sum_{k=l}^{q+1}\rho_k\left(1-\widehat{Y}_{l-1}(k;\boldsymbol{\rho})\right)}$, and $\widehat{h}_l^{v}(\boldsymbol{\rho}) = \frac{\sum_{k=0}^{l-1}\rho_k\Delta\widehat{Y}_l(k;\boldsymbol{\rho})}{\sum_{k=0}^{l-1}\rho_k\left(1-\widehat{Y}_{l-1}(k;\boldsymbol{\rho})\right)}$. In words, the hazard difference estimator quantifies directly avertible outcomes by multiplying the hazard difference $\widehat{h}_l^{u}(\boldsymbol{\rho}) - \widehat{h}_l^{v}(\boldsymbol{\rho})$ by the number survived among not-yet-vaccinated individuals $\widehat{N}_{l-1}^{u}(\boldsymbol{\rho})$.

**4 Bias of the hazard difference estimator relative to the causal estimand for vaccination at two time points**

**4.1 Analytic comparison**

Consider vaccination at two time points. The hazard difference estimator for quantifying outcomes that could have been directly averted under full vaccination at baseline $\boldsymbol{\psi} = \{2, d; 1{,}0\}$ but were not averted given the particular strategy $\boldsymbol{\rho} = \{2, d; \rho_0, \rho_1\}$ is:

$$\widehat{\delta_2}^{D*}(\boldsymbol{\psi},\boldsymbol{\rho}) = \widehat{N}_0^{u}(\boldsymbol{\rho})\left(\widehat{h}_1^{u}(\boldsymbol{\rho}) - \widehat{h}_1^{v}(\boldsymbol{\rho})\right) + \widehat{N}_1^{u}(\boldsymbol{\rho})\left(\widehat{h}_2^{u}(\boldsymbol{\rho}) - \widehat{h}_2^{v}(\boldsymbol{\rho})\right)$$

$$= N(\rho_1+\rho_2)\left(\frac{\rho_1\Delta\widehat{Y}_1(1;\boldsymbol{\rho}) + \rho_2\Delta\widehat{Y}_1(2;\boldsymbol{\rho})}{\rho_1+\rho_2} - \Delta\widehat{Y}_1(0;\boldsymbol{\rho})\right) + N\rho_2\left[1-\widehat{Y}_1(2;\boldsymbol{\rho})\right]$$

$$\cdot\left(\frac{\Delta\widehat{Y}_2(2;\boldsymbol{\rho})}{1-\widehat{Y}_1(2;\boldsymbol{\rho})} - \frac{\rho_0\Delta\widehat{Y}_2(0;\boldsymbol{\rho}) + \rho_1\Delta\widehat{Y}_2(1;\boldsymbol{\rho})}{\rho_0\left(1-\widehat{Y}_1(0;\boldsymbol{\rho})\right) + \rho_1\left(1-\widehat{Y}_1(1;\boldsymbol{\rho})\right)}\right)$$

$$= N\cdot\Bigg[(\rho_1+\rho_2)\left(\frac{\rho_1\Delta\widehat{Y}_1(1;\boldsymbol{\rho}) + \rho_2\Delta\widehat{Y}_1(2;\boldsymbol{\rho})}{\rho_1+\rho_2} - \Delta\widehat{Y}_1(0;\boldsymbol{\rho})\right) + \rho_2$$

$$\cdot\left(\Delta\widehat{Y}_2(2;\boldsymbol{\rho}) - \left[1-\widehat{Y}_1(2;\boldsymbol{\rho})\right]\cdot\frac{\rho_0\Delta\widehat{Y}_2(0;\boldsymbol{\rho}) + \rho_1\Delta\widehat{Y}_2(1;\boldsymbol{\rho})}{\rho_0\left(1-\widehat{Y}_1(0;\boldsymbol{\rho})\right) + \rho_1\left(1-\widehat{Y}_1(1;\boldsymbol{\rho})\right)}\right)\Bigg]$$

$$= N\cdot\Bigg[\left(\rho_1\Delta\widehat{Y}_1(1;\boldsymbol{\rho}) + \rho_2\Delta\widehat{Y}_1(2;\boldsymbol{\rho}) - (\rho_1+\rho_2)\cdot\Delta\widehat{Y}_1(0;\boldsymbol{\rho})\right) + \rho_2$$

$$\cdot\left(\Delta\widehat{Y}_2(2;\boldsymbol{\rho}) - \frac{(\rho_0+\rho_1)\cdot\left(1-\widehat{Y}_1(2;\boldsymbol{\rho})\right)}{\rho_0\left(1-\widehat{Y}_1(0;\boldsymbol{\rho})\right) + \rho_1\left(1-\widehat{Y}_1(1;\boldsymbol{\rho})\right)}\cdot\frac{\rho_0\Delta\widehat{Y}_2(0;\boldsymbol{\rho}) + \rho_1\Delta\widehat{Y}_2(1;\boldsymbol{\rho})}{\rho_0+\rho_1}\right)\Bigg].$$

Consider $E\left[\widehat{\delta_2}^{D*}(\boldsymbol{\psi},\boldsymbol{\rho})\right]$.

$$
\begin{aligned}
E\left[\widehat{\delta_2}^{D*}(\boldsymbol{\psi},\boldsymbol{\rho})\right] = E\Bigg[N \\
&\cdot \Bigg[\Big(\rho_1\Delta\hat{Y}_1(1;\boldsymbol{\rho}) + \rho_2\Delta\hat{Y}_1(2;\boldsymbol{\rho}) - (\rho_1+\rho_2)\cdot\Delta\hat{Y}_1(0;\boldsymbol{\rho})\Big) + \rho_2 \\
&\cdot \left(\Delta\hat{Y}_2(2;\boldsymbol{\rho}) - \frac{(\rho_0+\rho_1)\cdot\left(1-\hat{Y}_1(2;\boldsymbol{\rho})\right)}{\rho_0\left(1-\hat{Y}_1(0;\boldsymbol{\rho})\right)+\rho_1\left(1-\hat{Y}_1(1;\boldsymbol{\rho})\right)}\cdot\frac{\rho_0\Delta\hat{Y}_2(0;\boldsymbol{\rho})+\rho_1\Delta\hat{Y}_2(1;\boldsymbol{\rho})}{\rho_0+\rho_1}\right)\Bigg]\Bigg] \\
= N\cdot\Bigg\{&\left(\rho_1\Delta\bar{Y}_1(1;\boldsymbol{\rho}) + \rho_2\Delta\bar{Y}_1(2;\boldsymbol{\rho})\right) - (\rho_1+\rho_2)\cdot\Delta\bar{Y}_1(0;\boldsymbol{\rho}) + \rho_2\cdot\Delta\bar{Y}_2(2;\boldsymbol{\rho}) \\
&- E\left[\rho_2\cdot\left(\frac{(\rho_0+\rho_1)\cdot\left(1-\hat{Y}_1(2;\boldsymbol{\rho})\right)}{\rho_0\left(1-\hat{Y}_1(0;\boldsymbol{\rho})\right)+\rho_1\left(1-\hat{Y}_1(1;\boldsymbol{\rho})\right)}\cdot\frac{\rho_0\Delta\hat{Y}_2(0;\boldsymbol{\rho})+\rho_1\Delta\hat{Y}_2(1;\boldsymbol{\rho})}{\rho_0+\rho_1}\right)\right]\Bigg\} \qquad \text{(S24)}
\end{aligned}
$$

Comparing equation (S20) with equation (S24), we know that $E\left[\widehat{\delta_2}^{D*}(\boldsymbol{\psi},\boldsymbol{\rho})\right] \neq \delta_2^D(\boldsymbol{\psi},\boldsymbol{\rho})$ if vaccine has a non-null effect, as equation (S24) misses the term $\rho_1\cdot[\Delta\bar{Y}_2(1;\boldsymbol{\rho}) - \Delta\bar{Y}_2(0;\boldsymbol{\rho})]$ and $E\left[\rho_2\cdot\frac{(\rho_0+\rho_1)\cdot(1-\hat{Y}_1(2;\boldsymbol{\rho}))}{\rho_0(1-\hat{Y}_1(0;\boldsymbol{\rho}))+\rho_1(1-\hat{Y}_1(1;\boldsymbol{\rho}))}\cdot\frac{\rho_0\Delta\hat{Y}_2(0;\boldsymbol{\rho})+\rho_1\Delta\hat{Y}_2(1;\boldsymbol{\rho})}{\rho_0+\rho_1}\right] \neq \rho_2\cdot\Delta\bar{Y}_2(0;\boldsymbol{\rho})$.

**4.2 Simulation**

We simulate the epidemic under strategy $\boldsymbol{\rho}' = (2,60;0.2,0.3)$ based on the same model (equation [8]), scenarios (**Table 1**), parameters (**Table S4**), and initial conditions (**Table S5**) as with the simulations in the main text. We examine the bias of the hazard difference estimator $\widehat{\delta_2}^{D*}(\boldsymbol{\psi},\boldsymbol{\rho}')$ relative to the causal estimand $\delta_2^D(\boldsymbol{\psi},\boldsymbol{\rho}')$, where $\boldsymbol{\psi} = \{2,60;1,0\}$, and identify the conditions under which the bias would be substantial.

The hazard difference estimator overestimates directly avertible infections when $VE_{inf}$ = 90% (**Figure S6A; Table S13**). However, the hazard difference estimator overestimates directly avertible deaths in Scenarios 3 to 6 and underestimates directly avertible deaths in other scenarios (**Figure S6B; Table S13**).

Take Scenario 3 where IFR=100% as an example. The hazard difference estimator overestimates avertible deaths. Due to high $VE_{inf}$, survival is lower among $x = 2$ compared to the average of $x = 1$ or $x = 0$ (**Figure S3**). Specifically, $\frac{(\rho_0' + \rho_1') \cdot \left(1 - \hat{Y}_1(2; \boldsymbol{\rho}')\right)}{\rho_0'\left(1 - \hat{Y}_1(0; \boldsymbol{\rho}')\right) + \rho_1'\left(1 - \hat{Y}_1(1; \boldsymbol{\rho}')\right)} < 1$ in equation (S24), such that $E\left[\rho_2' \cdot \frac{(\rho_0' + \rho_1') \cdot \left(1 - \hat{Y}_1(2; \boldsymbol{\rho}')\right)}{\rho_0'\left(1 - \hat{Y}_1(0; \boldsymbol{\rho}')\right) + \rho_1'\left(1 - \hat{Y}_1(1; \boldsymbol{\rho}')\right)} \cdot \frac{\rho_0' \Delta\hat{Y}_2(0; \boldsymbol{\rho}) + \rho_1' \Delta\hat{Y}_2(1; \boldsymbol{\rho}')}{\rho_0' + \rho_1'}\right]$ in equation (S24) is smaller than $\rho_2' \cdot \Delta\bar{Y}_2(0; \boldsymbol{\rho}')$ in equation (S20).

However, in Scenario 2 where IFR=10%, the hazard difference estimator underestimates avertible deaths because the relative survival is similar (i.e., $\frac{(\rho_0' + \rho_1') \cdot \left(1 - \hat{Y}_1(2; \boldsymbol{\rho}')\right)}{\rho_0'\left(1 - \hat{Y}_1(0; \boldsymbol{\rho}')\right) + \rho_1'\left(1 - \hat{Y}_1(1; \boldsymbol{\rho}')\right)} \approx 1$), such that $E\left[\rho_2' \cdot \frac{(\rho_0' + \rho_1') \cdot \left(1 - \hat{Y}_1(2; \boldsymbol{\rho}')\right)}{\rho_0'\left(1 - \hat{Y}_1(0; \boldsymbol{\rho}')\right) + \rho_1'\left(1 - \hat{Y}_1(1; \boldsymbol{\rho}')\right)} \cdot \frac{\rho_0' \Delta\hat{Y}_2(0; \boldsymbol{\rho}') + \rho_1' \Delta\hat{Y}_2(1; \boldsymbol{\rho}')}{\rho_0' + \rho_1'}\right]$ in equation (S24) is greater than $\rho_2' \cdot \Delta\bar{Y}_2(0; \boldsymbol{\rho}')$ in equation (S20).

**TABLE S13** | Absolute and percentage bias of the hazard difference estimator relative to the causal estimand (reference) for avertible outcomes under strategy $\boldsymbol{\rho}' = (2{,}60; 0.2, 0.3)$ in scenarios varied by infection-fatality rate (IFR), vaccine efficacy against infection ($VE_{inf}$) and vaccine efficacy against death given infection ($VE_{death}$)

| | **Scenario 1** | **Scenario 2** | **Scenario 3** | **Scenario 4** | **Scenario 5** | **Scenario 6** | **Scenario 7** | **Scenario 8** | **Scenario 9** |
|---|---|---|---|---|---|---|---|---|---|
| | IFR: 1%, $VE_{inf}$ = 90%, $VE_{death}$ = 0% | IFR: 10%, $VE_{inf}$ = 90%, $VE_{death}$ = 0% | IFR: 100%, $VE_{inf}$ = 90%, $VE_{death}$ = 0% | IFR: 1%, $VE_{inf}$ = 0%, $VE_{death}$ = 90% | IFR: 10%, $VE_{inf}$ = 0%, $VE_{death}$ = 90% | IFR: 100%, $VE_{inf}$ = 0%, $VE_{death}$ = 90% | IFR: 1%, $VE_{inf}$ = 90%, $VE_{death}$ = 90% | IFR: 10%, $VE_{inf}$ = 90%, $VE_{death}$ = 90% | IFR: 100%, $VE_{inf}$ = 90%, $VE_{death}$ = 90% |
| **Avertible infections** [a] | 18348.79 (12.02%) | 19706.57 (12.92%) | 65014.35 (57.86%) | 0.00 | 0.00 | 0.00 | 18333.15 (12.01%) | 19528.3 (12.79%) | 46502.42 (35.06%) |
| **Avertible deaths** [a] | -300.49 (-19.75%) | -2911.27 (-19.12%) | 9526.54 (7.96%) | 0.11 (0.01%) | 11.35 (0.06%) | 1957.46 (1.01%) | -30.05 (-1.73%) | -291.24 (-1.67%) | -132.51 (-0.07%) |

[a] Percentage differences were calculated using values rounded to nine decimal places, to avoid uninterpretable, extremely small non-zero values (absolute value < $10^{-9}$). All values in the table are rounded to two decimal places.

**FIGURE S6** | Infections (A) and deaths (B) directly avertible among unvaccinated individuals with strategy $\boldsymbol{\rho}' = (2{,}60; 0.2, 0.3)$ under different scenarios varied by infection-fatality rate (IFR), vaccine efficacy against infection ($VE_{inf}$) and vaccine efficacy against death given infection ($VE_{death}$)

## 5 Bias of the hazard difference estimator relative to the causal estimand for vaccination at an arbitrary number of vaccination times

### 5.1 Analytic comparison

Now consider vaccination at an arbitrary number of vaccination times (i.e., $q \in \mathbb{N}$). First, expand $\delta_{q+1}^{D}(\boldsymbol{\psi}, \boldsymbol{\rho})$ in equation (S21) as:

$$\delta_{q+1}^{D}(\boldsymbol{\psi}, \boldsymbol{\rho}) = N \cdot \sum_{k=1}^{q+1} \rho_k \cdot \left(\bar{Y}_{q+1}(k; \boldsymbol{\rho}) - \bar{Y}_{q+1}(0; \boldsymbol{\rho})\right)$$

$$= N \cdot \sum_{k=1}^{q+1} \rho_k \cdot \left(\sum_{l=1}^{q+1} \Delta\bar{Y}_l(k; \boldsymbol{\rho}) - \sum_{l=1}^{q+1} \Delta\bar{Y}_l(0; \boldsymbol{\rho})\right)$$

$$= N \cdot \sum_{k=1}^{q+1} \rho_k \cdot \left(\Delta\bar{Y}_1(k; \boldsymbol{\rho}) + \sum_{l=2}^{q+1} \Delta\bar{Y}_l(k; \boldsymbol{\rho}) - \Delta\bar{Y}_1(0; \boldsymbol{\rho}) - \sum_{l=2}^{q+1} \Delta\bar{Y}_l(0; \boldsymbol{\rho})\right)$$

$$= N \cdot \left[\sum_{k=1}^{q+1} \rho_k \cdot \left(\Delta\bar{Y}_1(k; \boldsymbol{\rho}) - \Delta\bar{Y}_1(0; \boldsymbol{\rho})\right) + \sum_{k=1}^{q+1} \rho_k \cdot \left(\sum_{l=2}^{q+1} \Delta\bar{Y}_l(k; \boldsymbol{\rho}) - \sum_{l=2}^{q+1} \Delta\bar{Y}_l(0; \boldsymbol{\rho})\right)\right] \quad \text{(S25)}$$

Then, expand $\widehat{\delta_{q+1}}^{D*}(\boldsymbol{\psi}, \boldsymbol{\rho})$ in equation (S23) as:

$$\widehat{\delta_{q+1}}^{D*}(\boldsymbol{\psi}, \boldsymbol{\rho}) = \sum_{l=1}^{q+1} \left[\hat{N}_{l-1}^{u}(\boldsymbol{\rho}) \cdot \left(\hat{h}_l^{u}(\boldsymbol{\rho}) - \hat{h}_l^{v}(\boldsymbol{\rho})\right)\right]$$

$$= \sum_{l=1}^{q+1} \left\{ N \left[\sum_{k=l}^{q+1} \rho_k \left(1 - \hat{Y}_{l-1}(k; \boldsymbol{\rho})\right)\right] \cdot \left[\frac{\sum_{k=l}^{q+1} \rho_k \Delta\hat{Y}_l(k; \boldsymbol{\rho})}{\sum_{k=l}^{q+1} \rho_k \left(1 - \hat{Y}_{l-1}(k; \boldsymbol{\rho})\right)} - \frac{\sum_{k=0}^{l-1} \rho_k \Delta\hat{Y}_l(k; \boldsymbol{\rho})}{\sum_{k=0}^{l-1} \rho_k \left(1 - \hat{Y}_{l-1}(k; \boldsymbol{\rho})\right)}\right] \right\}$$

$$= N \sum_{l=1}^{q+1} \left\{ \sum_{k=l}^{q+1} \rho_k \Delta\hat{Y}_l(k; \boldsymbol{\rho}) - \left[\sum_{k=l}^{q+1} \rho_k \left(1 - \hat{Y}_{l-1}(k; \boldsymbol{\rho})\right)\right] \cdot \frac{\sum_{k=0}^{l-1} \rho_k \Delta\hat{Y}_l(k; \boldsymbol{\rho})}{\sum_{k=0}^{l-1} \rho_k \left(1 - \hat{Y}_{l-1}(k; \boldsymbol{\rho})\right)} \right\}$$

Pulling $l = 1$ from the summation →

$$= N \left\{ \sum_{k=1}^{q+1} \rho_k \Delta\hat{Y}_1(k; \boldsymbol{\rho}) - \left(\sum_{k=1}^{q+1} \rho_k\right) \cdot \Delta\hat{Y}_1(0; \boldsymbol{\rho}) \right.$$

$$\left. + \sum_{l=2}^{q+1} \left\{ \sum_{k=l}^{q+1} \rho_k \Delta\hat{Y}_l(k; \boldsymbol{\rho}) - \left[\sum_{k=l}^{q+1} \rho_k \left(1 - \hat{Y}_{l-1}(k; \boldsymbol{\rho})\right)\right] \cdot \frac{\sum_{k=0}^{l-1} \rho_k \Delta\hat{Y}_l(k; \boldsymbol{\rho})}{\sum_{k=0}^{l-1} \rho_k \left(1 - \hat{Y}_{l-1}(k; \boldsymbol{\rho})\right)} \right\} \right\}$$

$$= N\left\{\sum_{k=1}^{q+1}\rho_k\left(\Delta\hat{Y}_1(k;\boldsymbol{\rho})-\Delta\hat{Y}_1(0;\boldsymbol{\rho})\right)+\sum_{l=2}^{q+1}\left[\sum_{k=l}^{q+1}\rho_k\Delta\hat{Y}_l(k;\boldsymbol{\rho})-\left(\sum_{k=l}^{q+1}\rho_k\right)\cdot\frac{\frac{\sum_{k=l}^{q+1}\rho_k\left(1-\hat{Y}_{l-1}(k;\boldsymbol{\rho})\right)}{\sum_{k=l}^{q+1}\rho_k}}{\frac{\sum_{k=0}^{l-1}\rho_k\left(1-\hat{Y}_{l-1}(k;\boldsymbol{\rho})\right)}{\sum_{k=0}^{l-1}\rho_k}}\cdot\frac{\sum_{k=0}^{l-1}\rho_k\Delta\hat{Y}_l(k;\boldsymbol{\rho})}{\sum_{k=0}^{l-1}\rho_k}\right]\right\}.$$

Now consider $E\left[\widehat{\delta_{q+1}}^{D*}(\boldsymbol{\psi},\boldsymbol{\rho})\right]$.

$$E\left[\widehat{\delta_{q+1}}^{D*}(\boldsymbol{\psi},\boldsymbol{\rho})\right]=E\left[N\left\{\sum_{k=1}^{q+1}\rho_k\left(\Delta\hat{Y}_1(k;\boldsymbol{\rho})-\Delta\hat{Y}_1(0;\boldsymbol{\rho})\right)\right.\right.$$

$$\left.\left.+\sum_{l=2}^{q+1}\left[\sum_{k=l}^{q+1}\rho_k\Delta\hat{Y}_l(k;\boldsymbol{\rho})-\left(\sum_{k=l}^{q+1}\rho_k\right)\cdot\frac{\frac{\sum_{k=l}^{q+1}\rho_k\left(1-\hat{Y}_{l-1}(k;\boldsymbol{\rho})\right)}{\sum_{k=l}^{q+1}\rho_k}}{\frac{\sum_{k=0}^{l-1}\rho_k\left(1-\hat{Y}_{l-1}(k;\boldsymbol{\rho})\right)}{\sum_{k=0}^{l-1}\rho_k}}\cdot\frac{\sum_{k=0}^{l-1}\rho_k\Delta\hat{Y}_l(k;\boldsymbol{\rho})}{\sum_{k=0}^{l-1}\rho_k}\right]\right\}\right]$$

Using arguments parallel with those in the proof of Theorem S2 $\longrightarrow$

$$= N\cdot\left\{\left[\sum_{k=1}^{q+1}\rho_k\cdot\left(\Delta\bar{Y}_1(k;\boldsymbol{\rho})-\Delta\bar{Y}_1(0;\boldsymbol{\rho})\right)\right]\right.$$

$$\left.+\sum_{l=2}^{q+1}\left[\sum_{k=l}^{q+1}\rho_k\Delta\bar{Y}_l(k;\boldsymbol{\rho})\right]-E\left[\sum_{l=2}^{q+1}\left[\left(\sum_{k=l}^{q+1}\rho_k\right)\cdot\frac{\frac{\sum_{k=l}^{q+1}\rho_k\left(1-\hat{Y}_{l-1}(k;\boldsymbol{\rho})\right)}{\sum_{k=l}^{q+1}\rho_k}}{\frac{\sum_{k=0}^{l-1}\rho_k\left(1-\hat{Y}_{l-1}(k;\boldsymbol{\rho})\right)}{\sum_{k=0}^{l-1}\rho_k}}\cdot\frac{\sum_{k=0}^{l-1}\rho_k\Delta\hat{Y}_l(k;\boldsymbol{\rho})}{\sum_{k=0}^{l-1}\rho_k}\right]\right]\right\}\qquad\text{(S26)}$$

Equation (S25) and equation (S26) have the same first term. However, if vaccination has a non-null effect, $\sum_{k=1}^{q+1}\rho_k\cdot\left(\sum_{l=2}^{q+1}\Delta\bar{Y}_l(k;\boldsymbol{\rho})-\sum_{l=2}^{q+1}\Delta\bar{Y}_l(0;\boldsymbol{\rho})\right)=\sum_{k=1}^{q+1}\rho_k\cdot\sum_{l=2}^{q+1}\Delta\bar{Y}_l(k;\boldsymbol{\rho})-\sum_{k=1}^{q+1}\rho_k\cdot\sum_{l=2}^{q+1}\Delta\bar{Y}_l(0;\boldsymbol{\rho})$ in equation (S25) does not equal to $\sum_{l=2}^{q+1}\left[\sum_{k=l}^{q+1}\rho_k\Delta\bar{Y}_l(k;\boldsymbol{\rho})\right]-E\left[\sum_{l=2}^{q+1}\left[\left(\sum_{k=l}^{q+1}\rho_k\right)\cdot\frac{\frac{\sum_{k=l}^{q+1}\rho_k\left(1-\hat{Y}_{l-1}(k;\boldsymbol{\rho})\right)}{\sum_{k=l}^{q+1}\rho_k}}{\frac{\sum_{k=0}^{l-1}\rho_k\left(1-\hat{Y}_{l-1}(k;\boldsymbol{\rho})\right)}{\sum_{k=0}^{l-1}\rho_k}}\cdot\frac{\sum_{k=0}^{l-1}\rho_k\Delta\hat{Y}_l(k;\boldsymbol{\rho})}{\sum_{k=0}^{l-1}\rho_k}\right]\right]$ in equation (S26), implying $E\left[\widehat{\delta_{q+1}}^{D*}(\boldsymbol{\psi},\boldsymbol{\rho})\right]\neq\delta_{q+1}^{D}(\boldsymbol{\psi},\boldsymbol{\rho})$.

**5.2 Simulation**

We simulate the epidemic under strategy $\boldsymbol{\rho}'' = \{99{,}7; \mathbf{0.01}\}$ based on the model in equation (S14), scenarios in **Table 1**, initial values in **Table S2** and parameters in **Table S4**. Similar to findings observed under strategy $\boldsymbol{\rho}'$, the hazard difference estimator overestimates directly avertible infections when $VE_{inf}$=90% (**Figure S7A, Table S14**). It substantially overestimates directly avertible deaths when IFR=100% and $VE_{death}$ = 0% (Scenario 3), slightly overestimates them in Scenarios 4, 5, 6 and 9, and underestimates them in Scenarios 1, 2, 7, and 8 (**Figure S7B, Table S14**).

**TABLE S14** | Absolute and percentage bias of the hazard difference estimator relative to the causal estimand for avertible outcomes under strategy $\boldsymbol{\rho''} = \{99{,}7; \mathbf{0.01}\}$ in scenarios varied by infection-fatality rate (IFR), vaccine efficacy against infection ($VE_{inf}$) and vaccine efficacy against death given infection ($VE_{death}$)

| | **Scenario 1** | **Scenario 2** | **Scenario 3** | **Scenario 4** | **Scenario 5** | **Scenario 6** | **Scenario 7** | **Scenario 8** | **Scenario 9** |
|---|---|---|---|---|---|---|---|---|---|
| | IFR: 1%, $VE_{inf}$ = 90%, $VE_{death}$ = 0% | IFR: 10%, $VE_{inf}$ = 90%, $VE_{death}$ = 0% | IFR: 100%, $VE_{inf}$ = 90%, $VE_{death}$ = 0% | IFR: 1%, $VE_{inf}$ = 0%, $VE_{death}$ = 90% | IFR: 10%, $VE_{inf}$ = 0%, $VE_{death}$ = 90% | IFR: 100%, $VE_{inf}$ = 0%, $VE_{death}$ = 90% | IFR: 1%, $VE_{inf}$ = 90%, $VE_{death}$ = 90% | IFR: 10%, $VE_{inf}$ = 90%, $VE_{death}$ = 90% | IFR: 100%, $VE_{inf}$ = 90%, $VE_{death}$ = 90% |
| **Avertible infections** [a] | 48041.54 (25.08%) | 51050.02 (26.9%) | 183221.65 (295.78%) | 0.00 | 0.00 | 0.00 | 48034.29 (25.07%) | 50967.59 (26.85%) | 144266.72 (142.84%) |
| **Avertible deaths** [a] | -316.36 (-16.51%) | -2814.54 (-14.83%) | 133578.07 (215.64%) | 0.7 (0.03%) | 72.59 (0.3%) | 9981.36 (4.02%) | -31.53 (-1.22%) | -270.1 (-1.04%) | 10765.53 (4.17%) |

[a] Percentage differences were calculated using values rounded to nine decimal places, to avoid uninterpretable, extremely small non-zero values (absolute value $< 10^{-9}$). All values in the table are rounded to two decimal places

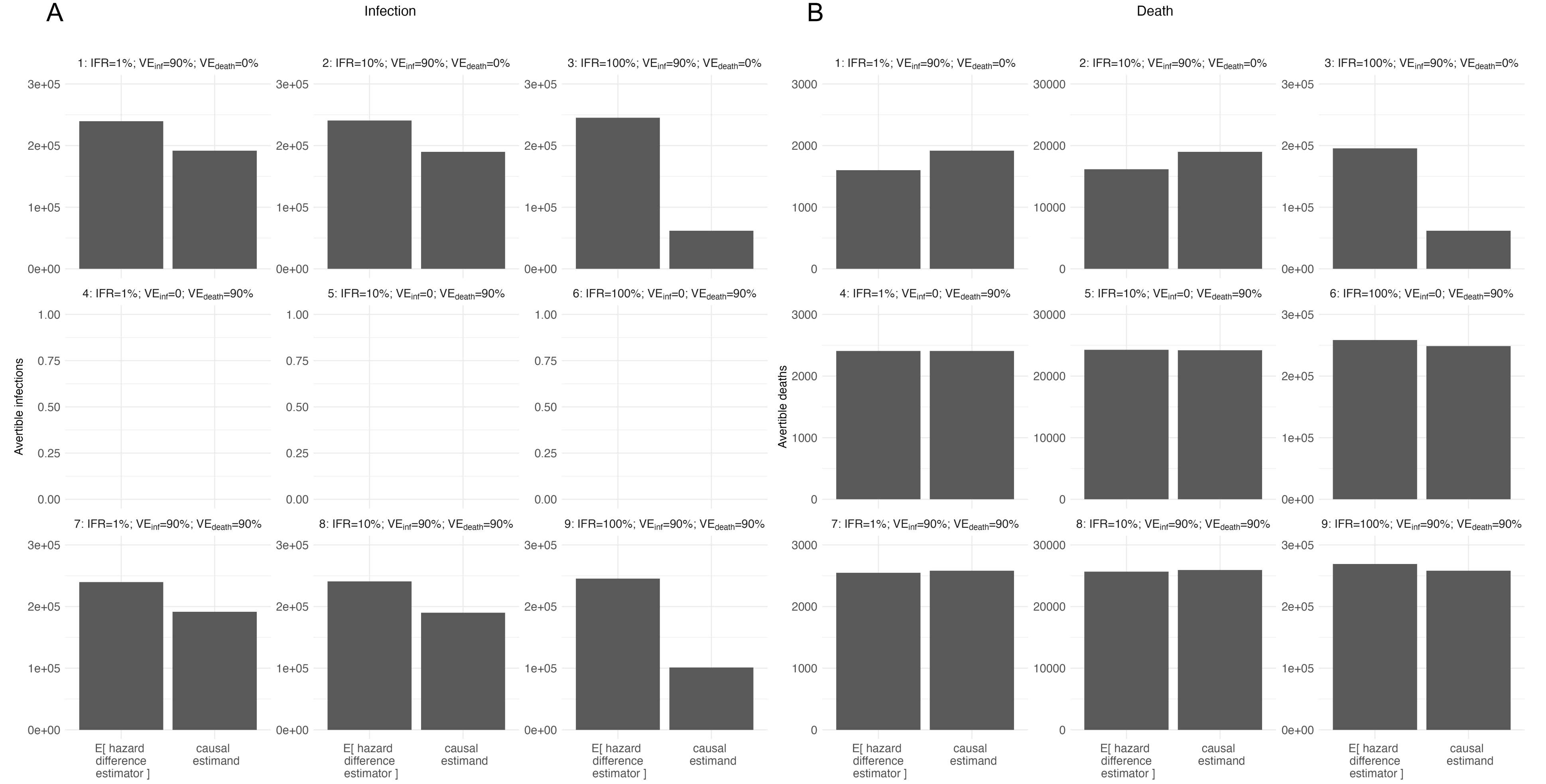

**FIGURE S7** | Infections (A) and deaths (B) directly avertible among unvaccinated individuals with strategy $\boldsymbol{\rho}'' = \{99{,}7; \mathbf{0.01}\}$ under different scenarios varied by infection-fatality rate (IFR), vaccine efficacy against infection ($VE_{inf}$) and vaccine efficacy against death given infection ($VE_{death}$)